\documentclass[journal=langd5,manuscript=article]{achemso}
\usepackage[version=3]{mhchem} % Formula subscripts using \ce{}
\usepackage{natbib}
\usepackage{url}
\usepackage{appendix}

\usepackage{epstopdf}
\usepackage{mathtools}
\usepackage{textgreek}

\usepackage{multirow}
\usepackage{multicol}
\usepackage{graphicx}
\usepackage{caption}

\usepackage{amssymb}
\usepackage{float}

\usepackage{color}

\author{Pallavi Katre}
\affiliation{Department of Chemical Engineering, Indian Institute of Technology Hyderabad, Sangareddy 502 284, Telangana, India}
\author{Sayak Banerjee}
\affiliation{Department of Mechanical and Aerospace Engineering, Indian Institute of Technology Hyderabad, Sangareddy 502 284, Telangana, India}
\author{Saravanan Balusamy}
\affiliation{Department of Mechanical and Aerospace Engineering, Indian Institute of Technology Hyderabad, Sangareddy 502 284, Telangana, India}
\author{Kirti Chandra Sahu}
\affiliation{Department of Chemical Engineering, Indian Institute of Technology Hyderabad, Sangareddy 502 284, Telangana, India}
\email{ksahu@che.iith.ac.in}

\title{Stability and retention force factor for binary-nanofluid sessile droplets on a inclined substrate}

\begin{document}

\maketitle

\begin{abstract}
We investigate the retention force factor of sessile droplets of pure (ethanol) and binary (water-ethanol) fluids laden with alumina nanoparticles placed on a critically inclined substrate. It is observed that while the critical angle of an ethanol droplet increases with an increase in nanoparticles concentration, for water-ethanol binary droplets, it reaches to plateau and decreases slightly after 0.6 wt.\% nanoparticle loading. The effect of composition and concentration of nanoparticles on the retention force factor is studied, and correlations are proposed for the retention force factor and critical angle for pure and binary droplets. Infrared images of evaporating droplets of pure and binary fluids reveal richer hydrothermal waves in droplets with nanoparticles loading than in droplets without loading, and these waves are more intense in pure ethanol droplets. On an inclined substrate, the body force caused the droplets to elongate more toward the receding side, which led to an earlier breakup of the droplet at the receding side. To the best of our knowledge, our study is a first attempt to investigate the retention force factor for the droplets loaded with nanoparticles on an inclined substrate.
\end{abstract}


\section{1. Introduction}
\label{sec:intro}
Evaporation of droplets is fascinating from a scientific and practical applications perspective due to its relevance in a wide range of applications, such as inkjet printing \cite{lim2012deposit, kim2006direct, park2006control, lim2009experimental, koo2006fabrication, tekin2004ink, de2004inkjet}, coating technology \cite{yanagisawa2014investigation}, combustion, hot-spot cooling, agriculture and microfluidics \cite{deegan1997capillary}, to name a few. In addition to the above-mentioned applications, droplet evaporation has applications in biological systems like fabrication of DNA microarrays \cite{dugas2005droplet,lee2006electrohydrodynamic}, determining the lifetime of respiratory droplets \cite{balusamy2021lifetime} and disease detection \cite{brutin2012influence, zeid2013influence, lanotte2017role}. The addition of nanoparticles to the base fluid (also called ``nanofluids") boosts the thermal conductivity and heat transfer rate. A nano-fluid droplet exhibits the ``coffee-ring" effect due to the deposition of nanoparticles near the contact line of the droplet. To examine the interesting physics, some researchers have examined the evaporation of droplets containing nanoparticles on horizontal substrates \cite{sefiane2014patterns,erbil2015control,zhong2014evaporation,zhong2016flow,parsa2017patterns,hari2022counter} and inclined substrates \cite{li2018pattern, mondal2018patterns,gopu2020evaporation}. The stability and retention force factor of pure fluid droplets on inclined substrates have also been investigated \cite{elsherbini2006retention,extrand1995liquid}. To the best of our knowledge, however, this is the first time we are reporting the droplet stability for pure fluid and binary fluid containing nanoparticles on inclined substrates.

First, we review the research on pure (single-component) droplets. Janardan and Panchagulla \cite{janardan2014effect} studied the shape of a sessile droplet on an inclined hysteretic substrate. The moving and sliding angles were calculated for different surfaces. They found that while the loss of global equilibrium and the onset of motion induces the sliding angle, the loss of local equilibrium causes the moving angle. The Bond number and the initial static contact angle were used to establish correlations between advancing and receding angles. The critical sliding angle and the sliding resistance of the droplet on a grooved surface were calculated by Ding et al. \cite{ding2020critical}. The geometric variations of a rolling droplet for different droplet sizes were investigated by Yilbas et al. \cite{yilbas2017dynamics}. They found that increasing the droplet size increases the drag, shear, and adhesion forces along the contact line. A scaling law was used to describe the dependency of sliding velocity on the inclination angle, and droplet volume \cite{kim2002sliding}. Droplet shape and wetting behaviour have been predicted for various inclination angles, and the critical inclination angle was found to be a function of droplet size \cite{annapragada2012droplet, chou2012drops}. The relationship between surface-tension forces and contact-angle hysteresis was also estimated using the retentive-force factor ($k$) \cite{elsherbini2006retention}. They found that the aspect ratio of a droplet has a negligible effect on the retention force. In both axisymmetric and asymmetric droplets, the initial width is almost constant for the same droplet volume, resulting in a steadily increasing $k$ with tilting \cite{rios2018effect}. For any solid/liquid combination and droplets with various shapes, a theory has been developed to calculate the retention force factor \cite{furmidge1962studies,extrand1995liquid}. It was found that the retention force factor values for pure droplets range from 1 to 3.14 \cite{elsherbini2006retention}.    

The majority of research on pure sessile droplets containing colloidal particles focused on the mechanism underlying deposition patterns. The sessile droplet dispersed with colloids produces a coffee-ring effect, whereas coffee-eyes are observed in the case of pendent droplets \cite{mondal2018patterns}. This is because the bulk flow advection in a pendant drop is directed toward the contact line, whereas interface-mediated transport is directed toward the apex of the drop. Gravity affects both the suspended particles and the droplet when it is placed on an inclined substrate, disrupting the symmetry of the particle deposition. The main causes of this asymmetric deposition include gravitational sedimentation, particle movement over the asymmetrically curved liquid-air interface, and droplet splitting at the very end of the evaporation process. Li et al. \cite{li2018pattern} investigated the impact of gravity on the deposition pattern of sessile and pendant droplets with sub-micron and micro-size particles. The morphology of particle deposition is governed by gravitational sedimentation, interfacial shrinkage, and outward capillary flow. The competition between gravitational sedimentation and interface shrinkage in the first stage decides whether the liquid-air interface can trap the particles. The second stage involves the capillary flow, which moves the particles towards the edge.

The evaporation and wetting dynamics of binary fluid droplets on horizontal and inclined surfaces were also studied by a few researchers \cite{ozturk2018evaporation,ozturk2020simple,li2019gravitational,diddens2017evaporating,sefiane2003experimental,sefiane2008wetting}. Yonemoto et al. \cite{yonemoto2018sliding} investigated the critical angle of inclination of the substrate as a function of surface energy density on a low-surface-energy solid for a water-ethanol binary mixture. The relationship between adhesion and gravitational force was analysed using a model with a particular contact area. According to Edwards et al. \cite{edwards2018density}, Gravitational force predominates the flow in the evaporation of microlitre-sized droplets. An equation which describes the motion of droplets due to gravitational, capillary, and Marangoni stresses resulting from the dependence of surface tension on local temperature was developed using lubrication theory on a heated inclined substrate. Mamalis et al. \cite{mamalis2016motion} considered a self-rewetting mixture of a binary fluid droplet and visualised the thermal patterns on the droplet placed on a heated inclined substrate. The presence of unique temperature patterns on evaporating droplets indicates the existence of thermocapillary/solutal effects owing to the internal flows. The interaction of the Marangoni stresses produced by the contact line caused it to move in the opposite direction to gravity.

A few researchers have also investigated the binary component droplets laden with nanoparticles. The evaporation of a binary fluid droplet of water and ethanol laden with graphite nanoparticles was investigated by Zhong, and Duan \cite{zhong2014evaporation}. It was observed that the evaporation behaviour deviates from the constant evaporation rate because the droplet containing more nanoparticles and ethanol evaporates more quickly. Increasing the concentration of nanoparticles increases the rate of evaporation. The droplet containing graphite nanoparticles has a higher pinning effect and a higher initial contact angle throughout the drying process than a pure water droplet. Three different flow regimes were observed during the evaporation process \cite{zhong2016flow}. The final deposition pattern was found to be the result of the relative weightings of stage 1 (when the nanoparticles migrate to the contact line) and stage 2 (when the Marangoni flow drives the nanoparticles to move inward). This behaviour is reinforced with an increasing load of ethanol. Parsa et al. \cite{parsa2017patterns} employed infrared thermography and optical microscopy to observe the three distinct flow patterns for water-ethanol binary droplets laden with copper oxide (CuO) at different substrate temperatures. On a non-heated substrate, a uniform deposition pattern was seen. However, on a heated substrate, dual rings and stick-slip were observed. The dynamics of the droplet as it evaporates were found to be similar to a water-butanol droplet without nanoparticles. As the gradient of surface tension decreases, convection currents and the chaotic motion of nanoparticles are also reduced. At room temperature, the difference in evaporation rate between a droplet containing a nanoparticle and one without nanoparticles is higher and diminishes as the substrate temperature rises. Katre et al. \cite{katre2021evaporation} investigated the evaporation dynamics of a water-ethanol binary mixture with alumina nanoparticles at different substrate temperatures. They observed that the droplet containing 0.6 wt.\% loading is pinned for most of its lifetime, whereas the wetting diameter of the droplet without loading decreases monotonically. The infrared images reveal that a droplet with nanofluid loading exhibits significantly richer thermal patterns than a droplet without nanoparticle loading. The droplet with nanoparticle loading shows vigorous mixing and a faster evaporation rate due to its pinning effect as well as thermo-capillary and thermo-solutal convection. The deposition pattern after the complete droplet evaporation shows that the nanoparticles were deposited near the triplet contact line, indicating the coffee ring effect. This symmetry is broken when the droplet is placed on an inclined substrate \cite{katre2022experimental}. In this case, the deposition of particles was more significant near the advancing side of the droplet, and an uneven stick-slip pattern was observed on the receding side.

The aforementioned review of the literature reveals that while many researchers have studied the evaporation dynamics, shapes, and deposition patterns of pure and binary droplets with and without nanoparticles, only a few studies have discussed the stability and retention force factor of droplets on an inclined substrate; albeit only for pure fluids and without nanoparticles. The droplet parameters and retention force play an important role in studying droplet stability at a critical inclination angle. In the present study, we investigate the stability and retention force factor for pure ethanol (E 100\% + W 0\%) and ethanol (E) and water (W) binary fluid droplets with alumina (Al$_2$O$_3$) nanoparticles of varying concentrations (wt.\%). In the present study, for binary fluid, we choose (E 80\% + W 20\%) composition based on our earlier investigations \cite{katre2021evaporation,katre2022experimental} for different compositions that show rich convection patterns for the (E 80\% + W 20\%) droplet. The correlations for the retention force factor and critical angle for pure and binary droplets are proposed. We also discuss the thermal patterns of the evaporating droplets by performing infrared thermography. The rest of the paper is organised as follows. The experimental setup and post-processing method are described in section 2. The results obtained from our experiments and the proposed correlations are discussed in section 3. Finally, we conclude the study in section 4.

\section{2. Experimental Methodology}
\label{sec:exp_methodology}

\subsection{2.1 Experimental set-up}

\begin{figure}[h]
\centering
\includegraphics[width=0.9\textwidth]{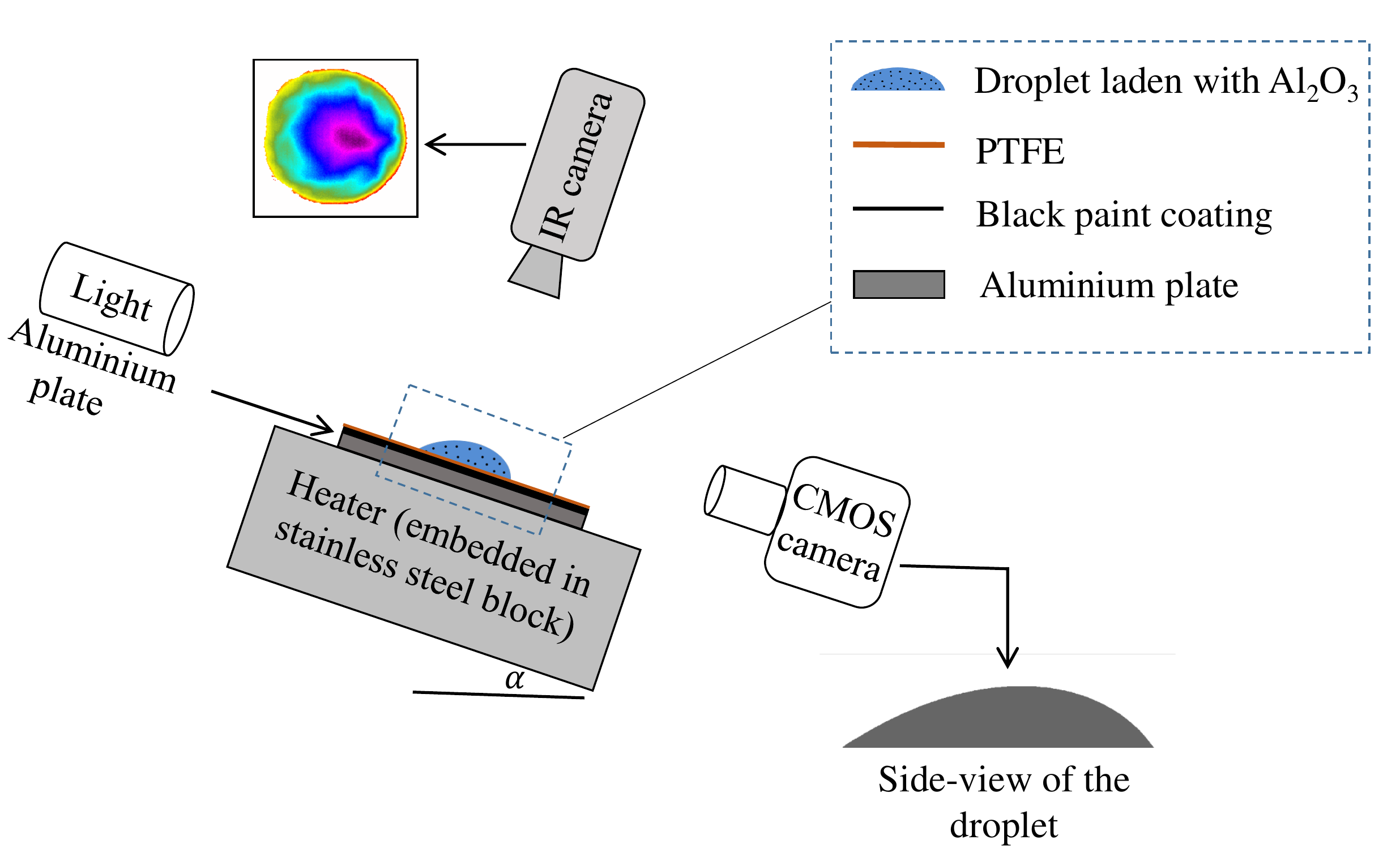}
\caption{Schematic of the experimental setup (customized goniometer). It consists of a heater placed in a stainless steel block, an aluminium plate with PTFE tape coated in black paint, a CMOS camera and light source arrangement and an infrared (IR) camera.}
\label{fig:fig1}
\end{figure}

The evaporation of sessile droplets of pure ethanol (E 100\% + W 0\%) and binary mixture (E 80\% + W 20\%) loaded with different concentrations (wt.\%) of alumina nanoparticles has been investigated using shadowgraphy and infrared (IR) imaging techniques. The experimental setup is shown schematically in figure \ref{fig:fig1}. We use a customised goniometer  (Make: Holmarc Opto-Mechatronic) for our experiments. It consists of a motorised pump with a syringe which produces droplets of required volume,  a multilayer metal block, a proportional-integral-derivative (PID)  controller which maintains the substrate temperature, an infrared (IR) camera (Make: FLIR, Model: X6540sc) and a metal-oxide-semiconductor (CMOS) camera (Model: DS-CBY501E-H). A light source is placed opposite to CMOS camera. The entire assembly was enclosed inside the goniometer box to minimise environmental disturbances from the outside. The goniometer box is maintained at 22$^\circ$C temperature and 50 $\pm$ 5\% relative humidity. The relative humidity is measured using a hygrometer (Make: HTC, Model: 288-ATH) that is installed inside the goniometer box.

The CMOS camera records the side view of the sessile droplet at a frame rate of 10 frames per second (fps) with a spatial resolution of $1280\times960$ pixels. The IR camera captured the top view of the droplet with a spectral range of 3 $\mu$m $-$ 5 $\mu$m, which displays the temperature profile on the liquid-air interface of the droplet. The thermal images are recorded at 50 fps with a spatial resolution of $640\times512$ pixels. The multilayer block consists of two electrical heaters controlled by a PID controller embedded in a stainless steel foundation and an aluminium plate of size 100 mm $\times$ 80 mm $\times$ 15 mm painted with black paint to minimise reflection in the infrared imaging system. We use a PTFE (polytetrafluoroethylene) tape having a thickness of 100$\mu$m as the substrate which is applied on a black painted aluminium plate. The stability of the PTFE tape is verified for the temperature range examined in this work. The roughness of the PTFE substrate measured using a digital microscope is found to be in between 4.74 and 18.16 $\mu$m \cite{katre2022experimental}. Prior to each experiment, the PTFE tape is cleaned with an isopropanol solution, dried using compressed air, and then placed on the metal plate. The substrate is maintained at a temperature $T_s = 50^\circ$C and is checked using a K-type thermocouple before placing the droplet on the substrate. To prepare the binary solution of (E 80\% + W 20\%), deionized water and absolute ethanol (99.9\% purity) are mixed with a stirrer to create a homogeneous mixture on a volume basis. The nanoparticles are then added by weight percentage (wt.\%). In our experiments, we use alumina (Al$_2$O$_3$) nanoparticles with a mean diameter of 20–30 nm purchased from Sisco Research Laboratories Pvt. Ltd. to prepare a mixture of various compositions laden with nanoparticles. To ensure homogeneous distribution in the solution, the mixture is ultrasonically treated for an hour (Make: BRANSONIC, CPX1800H-E). A motorised pump that regulates the volume flow rate was connected to a chromatographic syringe with a capacity of 100$\mu$m and a piston 8 size of 1.58 mm from Unitek Scientific Corporation. The syringe is fitted with a 21G needle with an aperture diameter of 0.51 mm which produces droplets with consistent size.

Pure and binary droplets of volume $(3 \pm 0.3)~\mu$l loaded with different nanoparticle concentrations are placed on a critically inclined substrate. The critical angle ($\alpha$) of inclination is the angle above which a droplet started to slide. It is to be noted that $\alpha$ depends on the composition of the fluid and the concentration of the nanoparticle. To find out the critical angle at a given condition, experiments are conducted with an initial inclination angle of 15$^\circ$ and with an increment of 5$^\circ$ until the droplet started to slide down at a particular angle. To get the exact value of the critical angle, an increment of $2^\circ$ is then used between the angles where the droplet slide and does not slide. 

\subsection{2.2 Post-processing}
The side view obtained from the CMOS camera is processed using an in-house developed \textsc{Matlab}$^{\circledR}$ program. The gradient is improved by utilising an unsharp masking approach to sharpen the image and a median filtering technique to remove random noise. After being filtered, the image is transformed into a binary image using an appropriate threshold that distinguishes the droplet boundary from the surrounding area. The reflection due to light is then eliminated, and the droplet contour was traced using a \textsc{Matlab}$^{\circledR}$ tool. The detailed post-processing method can be found in our previous study \cite{gurrala2019evaporation}. To process the infrared images, the droplet contour is extracted using the U-net machine learning model as discussed in our previous work \cite{katre2021evaporation}. It is usual to practice edge detection and intensity thresholding to separate the droplet contour from the background. Additionally, the U-net architecture uses data augmentation by elastically deforming the annotated input photos, which enables the network to make greater use of the available annotated images. The U-net-based machine learning model only requires a few annotated images. Therefore a total of 40 manually annotated grey-scale infrared images are used to train the network on a computer with a GPU (NVIDIA Quadro P1000). The network extracts the binary masks and droplet boundaries from the infrared images. Finally, the background is removed using a \textsc{Matlab}$^{\circledR}$ function, and the thermal profiles of the evaporating droplets are analysed.

\section{3. Results and discussion}
\label{sec:results}

We investigate the droplet stability and retention force factor of pure ethanol (E 100\% + W 0\%) and binary (E 80\% + W 20\%) sessile droplets with and without Al$_2$O$_3$ nanoparticles loading at the onset of sliding. The substrate temperature is maintained at 50$^\circ$C. The concentration of nanoparticles (in wt.\%) in the solution is varied from 0 wt.\% to 1 wt.\%, and its effect on the retention force factor is studied. 
 
\begin{figure}[h]
\centering
\includegraphics[width=1.0\textwidth]{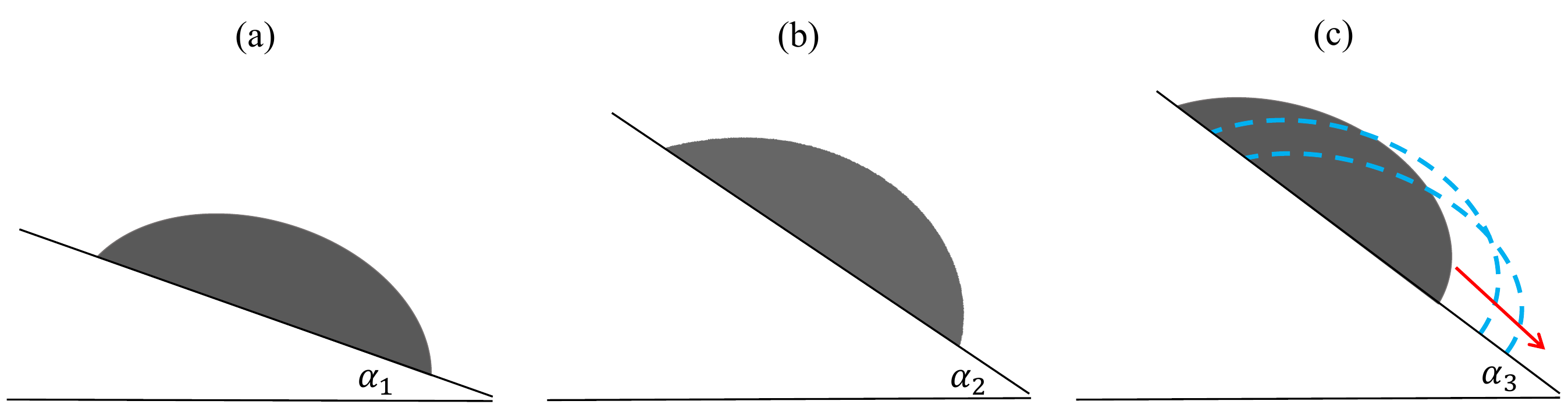}
\caption{Side-view of a (E 80\% + W 20\%) droplet with 0.6 wt.\% nanoparticles loading placed on a substrate with different inclination angle, $\alpha$. (a) $\alpha = 20^\circ \pm 1^\circ$ (almost symmetrical droplet), (b) $\alpha = 35^\circ \pm 1^\circ$ (a droplet at its critical inclination angle) and (c) $\alpha = 37^\circ \pm 1^\circ$ (a droplet that starts to slide). In panel (c), the dashed lines represent the profiles of the sliding droplet.}
\label{fig:fig2}
\end{figure}

A sessile droplet starts to slide when the inclination angle exceeds the critical angle, while the surface tension force allows the droplet to stick to the inclined or vertical substrate. The two forces that act on a sessile droplet placed on an inclined substrate are (i) surface tension force ($F_s$) and (ii) gravitational force. On a critically inclined surface, $F_s = mg\sin\alpha$ (which is the tangential component of gravitational force). When the droplet is placed on an inclined substrate, the front edge of the drop moves forward, whereas the rear edge remains fixed. As this happens, the advancing angle increases, and the receding angle decreases. Figure \ref{fig:fig2} shows the different profiles of an (E 80\% + W 20\%)  droplet loaded with 0.6 wt.\% nanoparticles with changing inclination angles of the substrate. It can be seen that when the droplet is placed at a lower inclination angle ($\alpha = 20^\circ\pm1^\circ$) than its critical angle, the surface tension force dominates the gravitational force, and the droplet is stable (figure \ref{fig:fig2}a). Figure \ref{fig:fig2}b depicts a droplet placed at its critical angle ($\alpha = 35^\circ\pm1^\circ$). The critical angle corresponds to the situation when the surface tension and gravitation forces balance each other. A further increase in the inclination angle causes the droplet to slide. Figure \ref{fig:fig2}c shows a droplet placed above its critical inclination angle, and the blue dashed lines show profiles of the sliding droplet. In this case, the droplet starts to slide as a parallel component of gravity ($mg\sin\alpha$) overcomes the surface tension force. The critically inclined droplets having non-spherical contours create an effective surface tension force upward along the inclination, counteracting the gravitational force component. Because of this, it is simple to calculate the upward component of the surface tension force using the contact angle hysteresis data. The relationship between surface tension force ($F_s$) holding a droplet on a solid substrate and the contact angles is given by \cite{elsherbini2006retention}
\begin{equation}
   \frac{F_s}{\gamma R} = k (\cos \theta_R - \cos \theta_A),
\label{eq:eq1}
\end{equation}
where $\theta_A$ and $\theta_R$ are advancing and receding contact angles, respectively, $\gamma$ is liquid-gas surface tension, and $R$ is the length scale that reflects the droplet contour's size. The advancing ($\theta_A$) and receding contact angles ($\theta_R$) of a droplet are depicted in supplementary figure S1. The values of $\theta_A$ and $\theta_R$ for (E 100\% + W 0\%) and (E 80\% + W 20\%) droplets placed at their respective critical angles for different values of nanoparticle loading (wt.\%) are presented in supplementary table ST1.  In the present study, due to the non-spherical contour of the droplet, $R$ is calculated as $(D_1 + D_2)/2$, wherein $D_1$ and $D_2$ are the diameters of the droplet along and across the inclination direction, respectively. In Eq. (\ref{eq:eq1}), $k$ is a constant called as retention force factor. It is to be noted that $k$ depends on the droplet geometry \cite{extrand1995liquid}. For pure fluid droplets, the retention force factor is between 1 and 3.14 \cite{elsherbini2006retention}. However, no one has yet reported the retention force factor for binary-nanofluid droplets, which is the objective of the present work. Thus it is possible to directly evaluate the force required to cause any droplet to move with any contact angle hysteresis if the values of $R$ and $k$ are known. We have restricted our investigation in the present study to Newtonian fluids only. Other classes of fluids may exhibit a different retention force factor.

In the present study, we investigate the effect of different parameters like concentration of nanoparticles (in wt.\%) and composition on $k$ values for sessile droplets placed on their respective critically inclined substrates. Supplementary table ST2 presents the critical angle values for different wt.\% of nanoparticle for (E 100\% + W 0\%)  and (E 80\% + W 20\%) droplets. It is observed that the critical angle of the (E 100\% + W 0\%) droplet is less than the (E 80\% + W 20\%) droplet for no loading (0 wt.\%) condition as water has a higher surface tension than ethanol. The critical angle increases with an increase in the nanoparticle wt.\% for the (E 100\% + W 0\%) droplet because the surface tension of the droplet increases with the addition of nanoparticles \cite{olayiwola2019mathematical} and the upward component of this surface tension force can balance with the downward gravitational component even at higher inclination angles. However, in the case of the (E 80\% + W 20\%) droplet, the critical angle reaches a plateau value at about 0.6 wt.\% loading and even decreases with a further increase in nanoparticle wt.\%. This may be because the further addition of nanoparticles increases the weight of the droplet, which leads to an increase in the downward component of gravity. This is not fully compensated by the increase in the surface tension force due to the addition of nanoparticles. This does not apply to the (E 100\% + W 0\%) droplet as the (E 80\% + W 20\%) droplet has a higher weight due to the presence of water (which is heavier than ethanol).

% \begin{table}[h]
% \centering
% \caption{The values of the critical angle of inclination ($\alpha$, in degree) of (E 100\% + W 0\%) and (E 80\% + W 20\%) droplets at the onset of sliding for different values of nanoparticle loading (wt.\%).}
% \label{table:T1}
% \begin{tabular}{|c|c|c|c|c|c|c|} 	\hline
%  \backslashbox[0.51cm]{Composition}{wt.\%}  & 0     & 0.2  & 0.4  & 0.6   & 0.8    & 1.0 
%      \\ \hline
% E 100\% + W 0\% &  $18 \pm 1$    & $25 \pm 1$  & $31 \pm 1$  & $35 \pm 1$   & $36 \pm 1$   & $37 \pm 1$ \\ \hline
% E 80\% + W 20\%  & $20 \pm 1$     & $25\pm 1$   & $31 \pm 1$  & $35 \pm 1$   & $30 \pm 1$    & $29 \pm 1$ 
%      \\ \hline

% \end{tabular}
% \end{table}

\subsection{3.1 Droplet profile: side view}

\begin{figure}[h]
\centering
\includegraphics[width=0.9\textwidth]{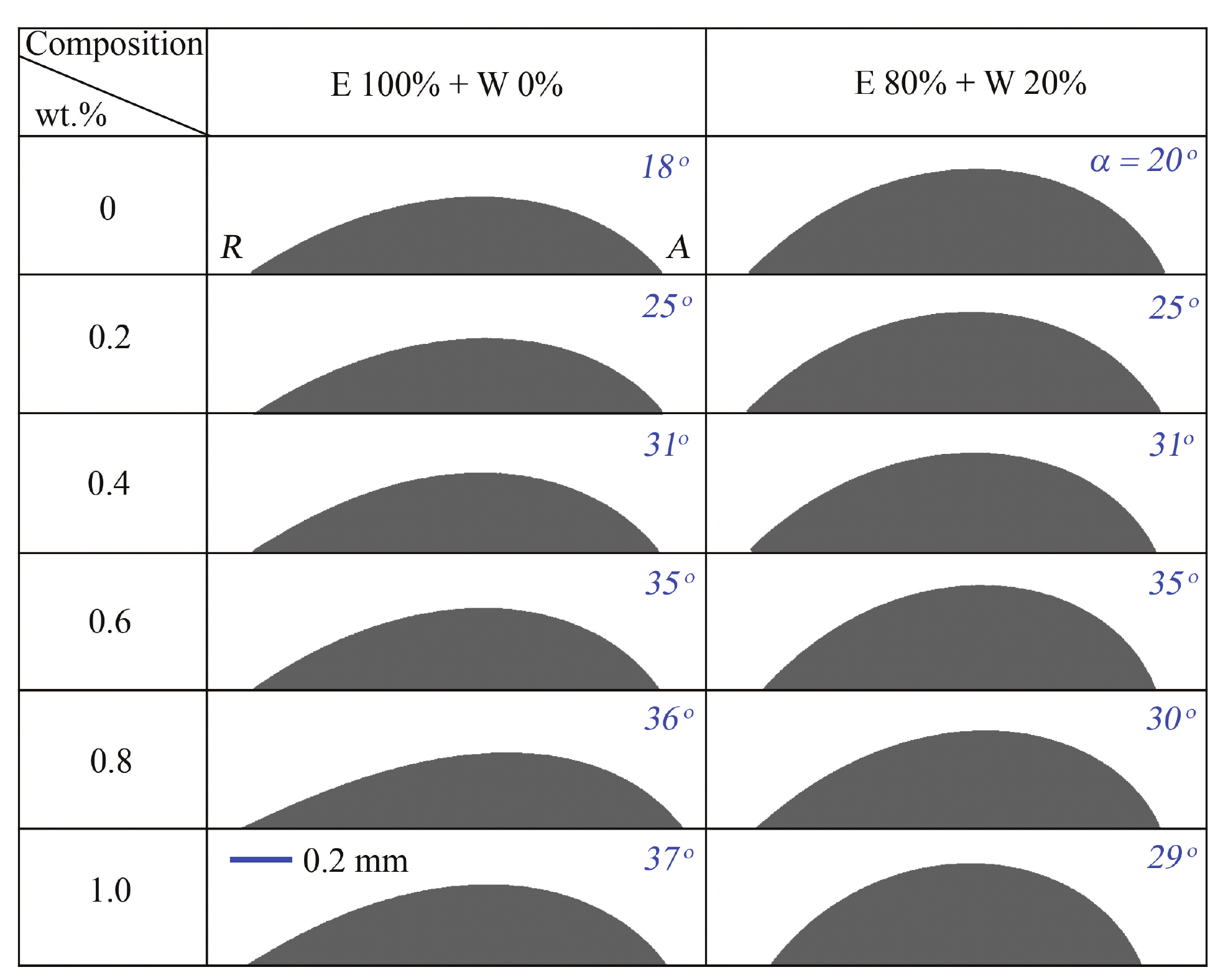}
\caption{Side-view of (E 100\% + W 0\%) and (E 80\% + W 20\%) droplets with 0.6 wt.\% nanoparticle loading placed on a substrate inclined at the corresponding critical inclination angle. The values of the critical angle ($\alpha$) are included at the top-right corner of each panel.}
\label{fig:fig3}
\end{figure}

\begin{figure}[h]
 \centering
 \hspace{0.6cm}  (a) \hspace{7.2cm} (b) \\
\includegraphics[width=0.45\textwidth]{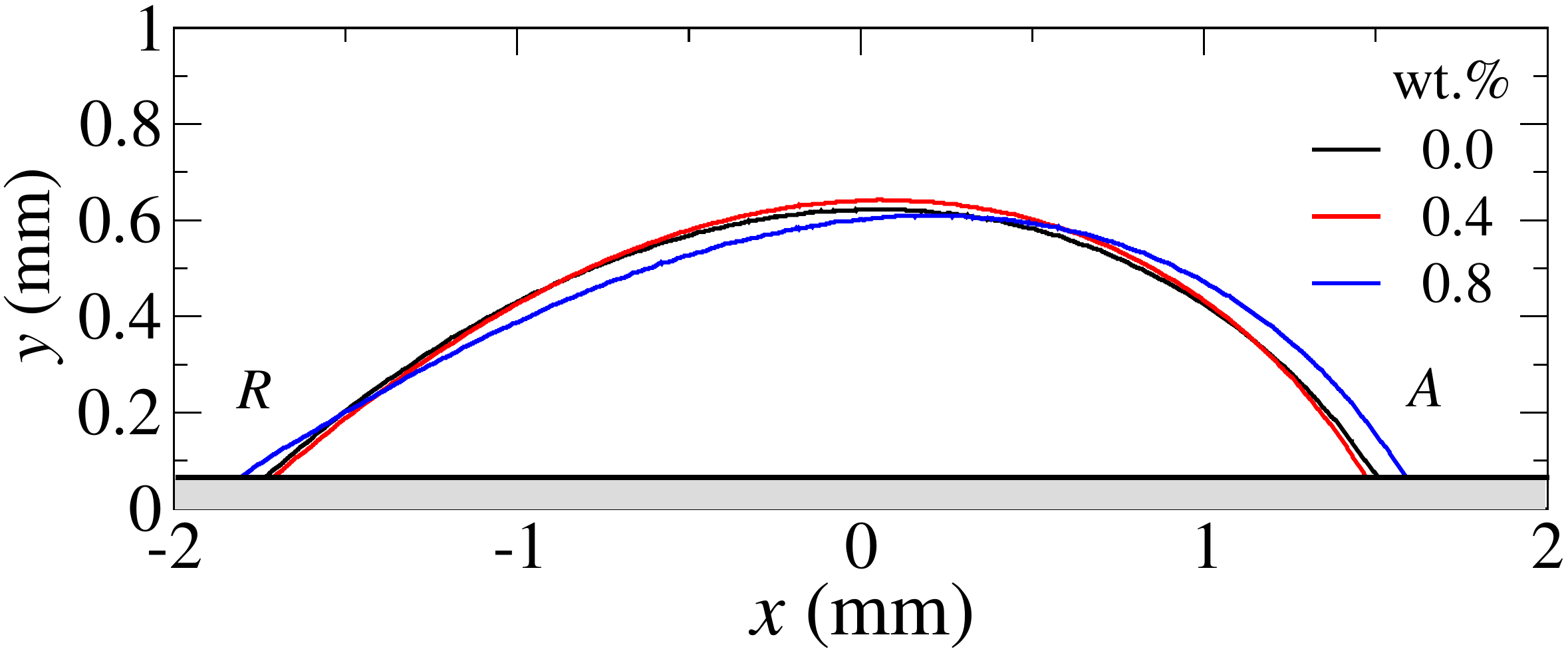} \hspace{2mm} \includegraphics[width=0.45\textwidth]{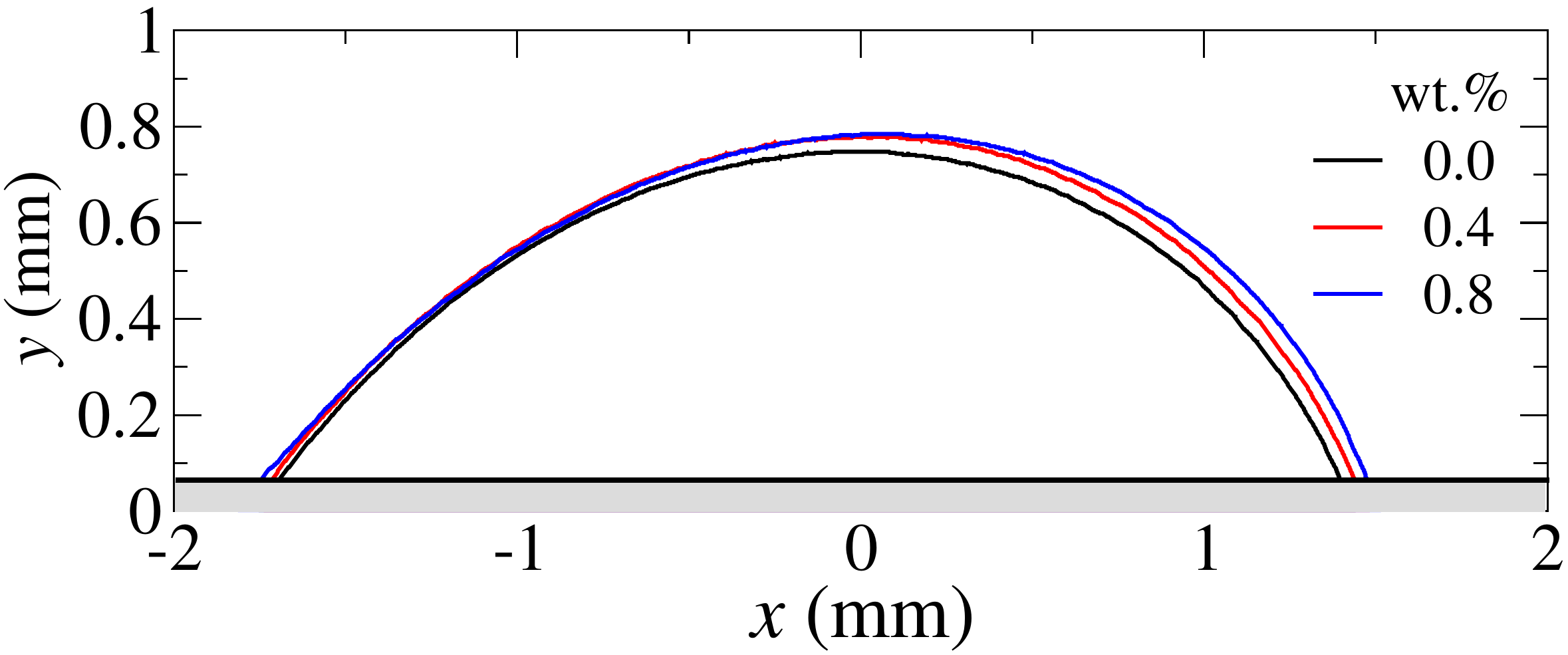}
\caption{Comparison of the droplet profile with different values of the nanaoparticles loading (wt.\%) when it is placed on a substrate inclined at the corresponding critical inclination angle. The panels (a) and (b) are associated with (E 100\% + W 0\%) and (E 80\% + W 20\%) droplets, respectively.}
\label{fig:fig4}
\end{figure}

This section analyses the initial side view profiles of (E 100\% + W 0\%) and (E 80\% + W 20\%) droplets laden with different loading conditions (wt.\%) placed at their respective critical angles (figure\ref{fig:fig3}). The initial volume of the droplet was maintained practically constant by performing many repetitions. In figure\ref{fig:fig3}, $R$ and $A$ indicate the receding and advancing side, respectively. As more liquid is shifted toward the advancing side due to body force, the contact angle on the advancing side is greater than the receding side for all cases. Due to its lower surface tension, the (E 100\% + W 0\%) droplet spreads more than the (E 80\% + W 20\%) droplet and exhibits lower contact angle values. It can be seen that (E 100\% + W 0\%) droplets exhibit similar profiles for small nanoparticle loadings (wt.\% $\le 0.6$). However, a noticeable difference is observed in the profiles of droplets containing 0.8 wt.\% and 1.0 wt.\%. The spread of the droplet with 0.8 wt.\% loading is higher and exhibits lower contact angles than that of the droplet with 1.0 wt.\% loading. This may be because a droplet with 1.0 wt.\% loading deposits more nanoparticles near the triple contact line than a droplet with 0.8 wt.\% loading. Similar behavior is shown for the (E 80\% + W 20\%) droplet with 1.0 wt.\%, which exhibits a smaller spread compensated by higher height and contact angles. To illustrate this behaviour, in figure \ref{fig:fig4}(a) and (b), the superimposed contours profiles of the droplets with different wt.\% of nanoparticles loading are plotted for (E 100\% + W 0\%) and (E 80\% + W 20\%) droplets, respectively. Here, the substrate is maintained at 50$^\circ$C.

The values of critical angle correspond to (E 100\% + W 0\%) and (E 80\% + W 20\%) droplets for different loadings (wt.\%) are presented graphically in figure \ref{fig:fig5}(a) (also see, table ST2 of the supplementary material). It can be seen that, for both (E 100\% + W 0\%) and (E 80\% + W 20\%) droplets, increasing the nanoparticles loading (wt.\%) up to 0.6 wt.\% monotonically increases the critical angle $(\alpha)$. Then the behaviour diverges for the two cases. For (E 80\% + W 20\%) droplet, the critical angle is highest for 0.6 wt.\% and starts an appreciable decline up to 1 wt.\% loading. On the other hand, for (E 100\% + W 0\%) droplet, the critical angle nearly stabilises after 0.6 wt.\%. The variations of the Bond number, $Bo$, that signifies the competition between the effective body force acting on the droplet with the surface tension force as defined in Eq. (\ref{eq:eq2}) with nanoparticle loading (wt.\%) for (E 100\% + W 0\%) and (E 80\% + W 20\%) droplets are presented in figure \ref{fig:fig5}(b).

The Bond number, $Bo$ is given by
\begin{equation}
Bo =  {4 \rho_{\rm eff} g R^2 \sin \alpha \over \gamma}.
\label{eq:eq2}
\end{equation}
where $g$ is acceleration due to gravity and $\rho_{\rm eff}$ is the effective density of the fluid. For the droplet laden with nanoparticles, $\rho_{\rm eff}$ is calculated using mixture rule as 
\begin{equation}
\rho_{\rm eff}(t) = (1 - Y_n (t)) \rho_f(t) + Y_n (t) \rho_n.
\label{eq:eq3}
\end{equation}
Here, $\rho_f$ and $\rho_n$  are the densities of the base fluid and base fluid laden with nanoparticles, respectively. The surface tension for the ethanol-water binary mixture at the substrate temperature is taken from the literature \cite{vazquez1995surface}. Previous studies have shown that the surface tension of liquids with tiny quantities of nanoparticles does not significantly deviate from the value of the surface tension of pure liquids \cite{tanvir2012surface}.

Figure \ref{fig:fig5}b shows that for the (E 100\% + W 0\%) droplet, the Bond number increases with an increase in nanoparticle wt.\% up to 0.8 wt.\% and then declines. In Eq. (\ref{eq:eq2}), it can be observed that the Bond number is directly proportional to $\sin \alpha$, $R^2$ and $\rho_{eff}$. Thus, it can be inferred that the increase in the Bond number up to 0.8 wt.\% is caused by the rise in the critical angle and $\rho_{eff}$ of the droplet. It can be seen that for 1.0 wt.\%, there is a rise in the critical angle by 1$^\circ$. The influence of $R^2$ prevails on the Bond number since $R$ is lower for 1.0 wt.\% than for 0.8 wt.\% (see, figure \ref{fig:fig3}), causing a decrease in the Bond number. For (E 80\% + W 20\%) droplet, the Bond number rises till 0.6 wt.\%, decreases for 0.8 wt.\%, and then slightly rises for 1.0 wt.\%. The value of the Bond number changes when we increase the nanoparticle loading because of an increase in the effective density and change in the critical angle of the droplet. The rate of increase or decrease of Bond number won't track the change in critical angle exactly. When the change in critical angle is modest, the wt.\% also plays a role. When wt.\% changes from 0.6 wt.\% to 0.8 wt.\%, there is a decrease in the critical angle by $5^\circ$, and the Bond number also decreases. However, the critical angle changes a little ($1^\circ$) from 0.8 wt.\% to 1.0 wt.\%. As a result, the rise in the nanoparticles loading also comes into play, and we notice a slight divergence in the response of the Bond number.

\begin{figure}[h]
 \centering
 \hspace{0.6cm}  (a) \hspace{7.2cm} (b) \\
  \includegraphics[width=0.45\textwidth]{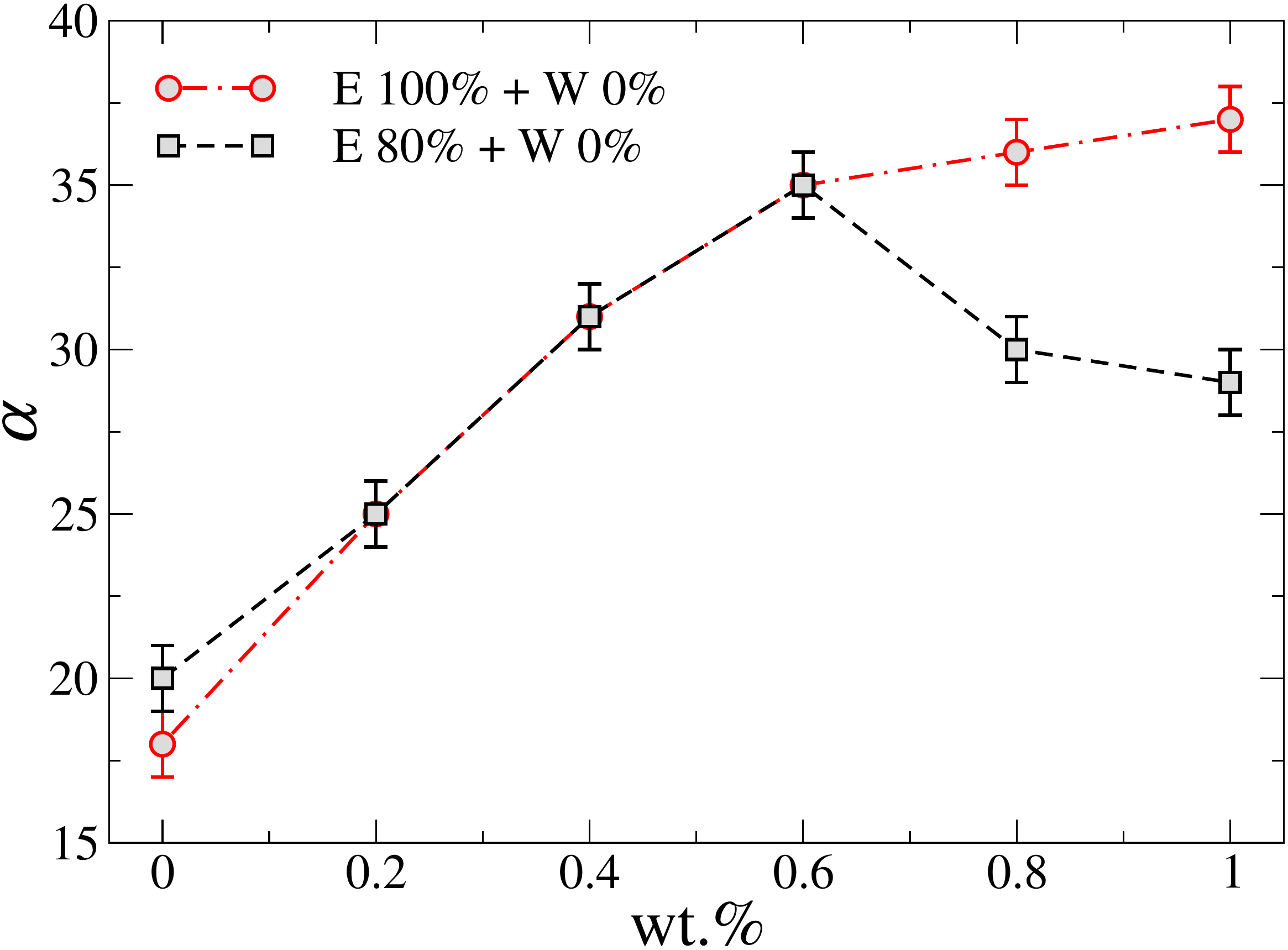} \hspace{2mm} \includegraphics[width=0.45\textwidth]{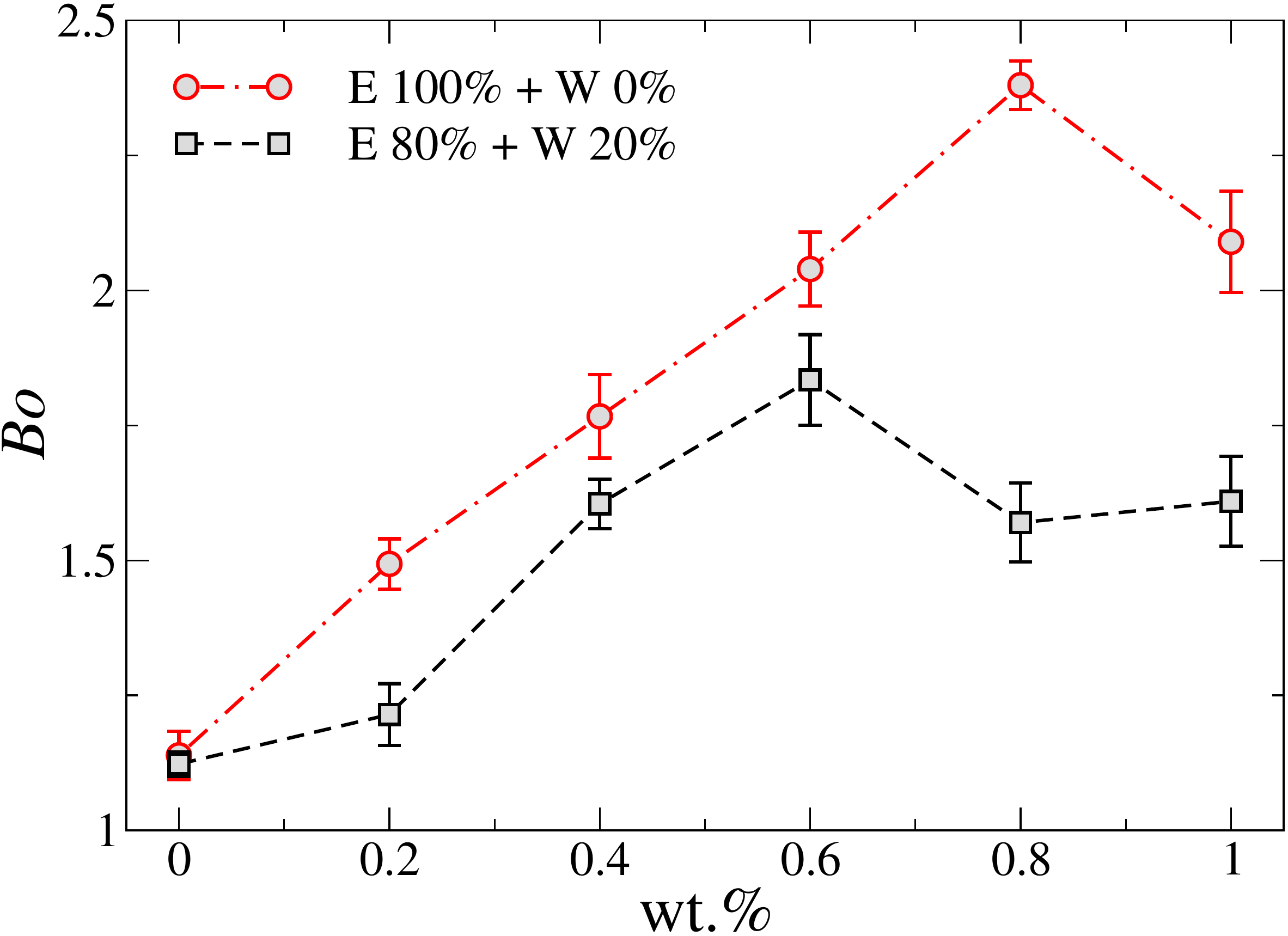}\\
\caption{Variation of the critical inclination angle ($\alpha$) and the Bond number ($Bo$) with different nanoparticle loadings (wt.\%) for (E 100\% + W 0\%) and (E 80\% + W 20\%) droplet.}
\label{fig:fig5}
\end{figure}

\subsection{3.2 Contact angle hysteresis}

\begin{figure}[h]
\centering
\hspace{0.6cm}  (a) \hspace{7.3cm} (b) \\
\includegraphics[width=0.45\textwidth]{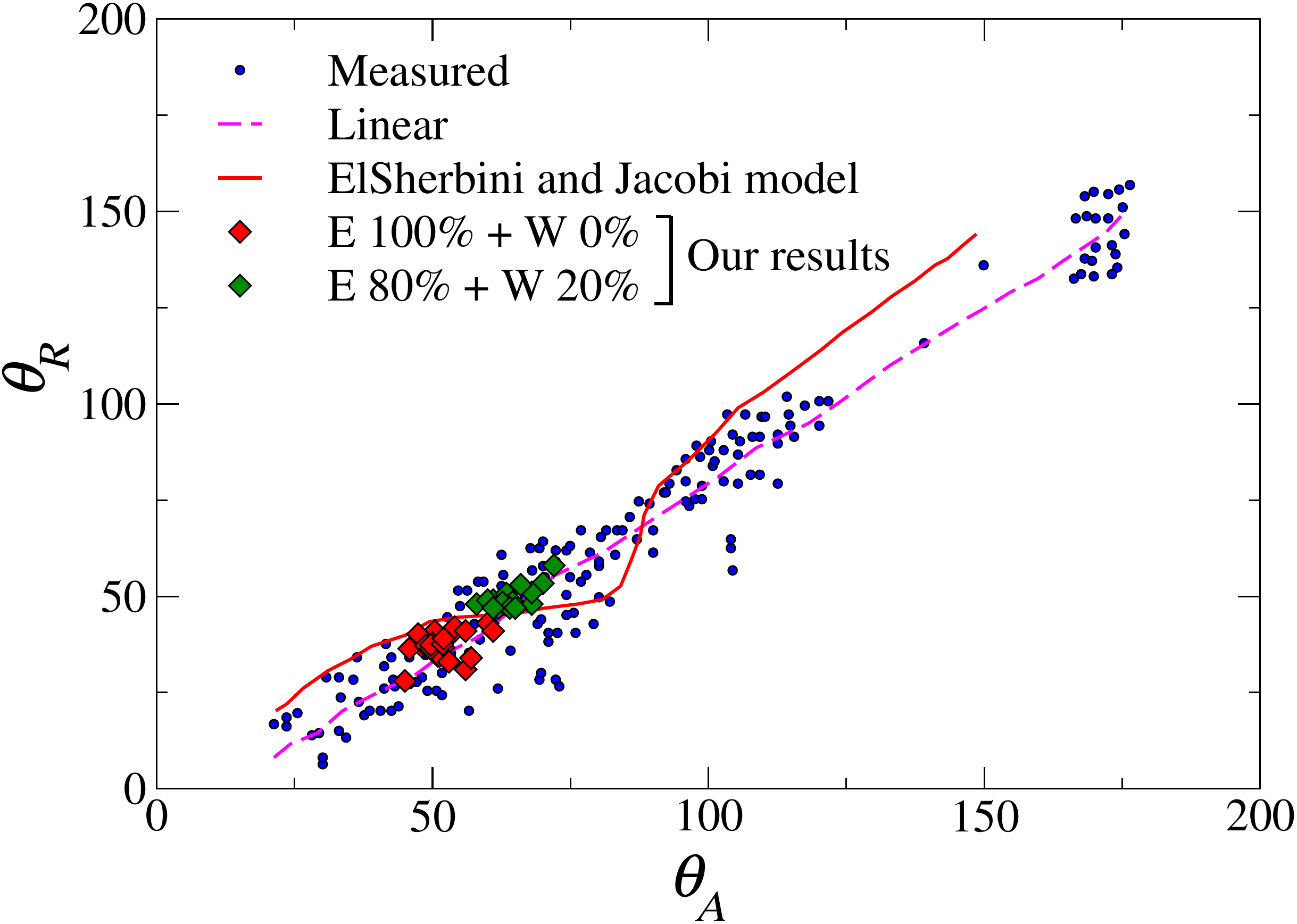} \hspace{2mm} \includegraphics[width=0.45\textwidth]{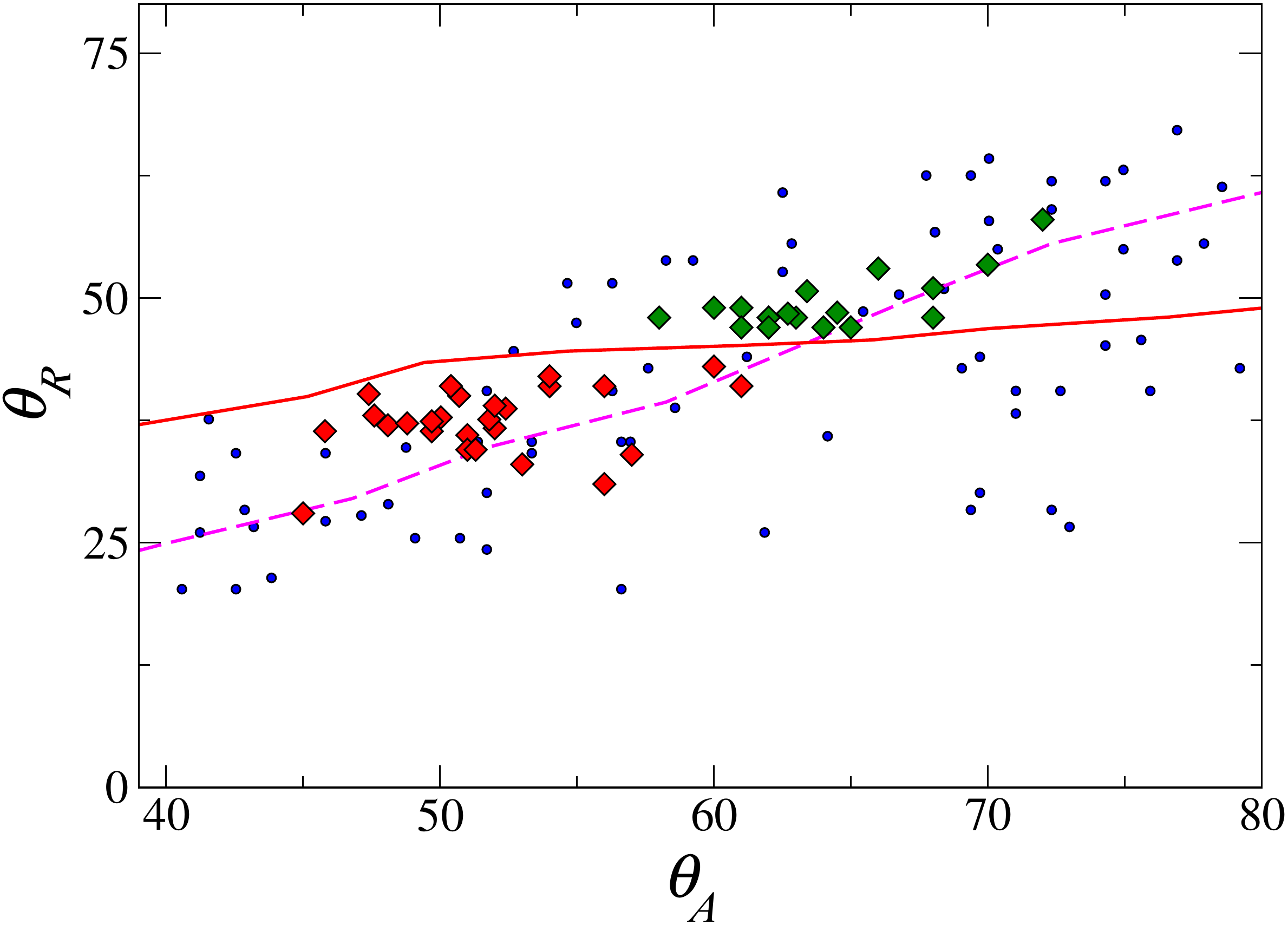}\\
\caption{(a) Comparison of the experimentally obtained advancing ($\theta_A$) and receding ($\theta_R$) contact angles with that obtained from the model and linear fit. The dashed magenta line shows the linear fit, the solid red line is associated with the model developed by ElSherbini and Jacobi \cite{elsherbini2006retention} and blue dots show the measurements from different sources \cite{macdougall1942surface, goodwin1988model, quere1998drops, extrand1997experimental, chibowski2002interpretation, extrand2002water, schmidt2004contact, faibish2002contact}. (b) A magnified view of panel (a) to show our results represented by red and green diamond symbols for (E 100\% + W 0\%) and (E 80\% + W 20\%) droplets, respectively.}
\label{fig:fig6}
\end{figure}

The advancing ($\theta_A$) and receding contact angle ($\theta_R$) obtained from our experiments are plotted in figure \ref{fig:fig6}(a) and (b) along with the results from different sources \cite{macdougall1942surface, goodwin1988model, quere1998drops, extrand1997experimental, chibowski2002interpretation, extrand2002water, schmidt2004contact, faibish2002contact}. They provided the values of $\theta_A$ and $\theta_R$ for a variety of liquids and surfaces, including various liquid compositions and surface conditions as presented by blue dots. ElSherbini and Jacobi's model \cite{elsherbini2006retention} based on the two-circle analysis fits this data with a correlation coefficient of 0.90. A linear fit ($\theta_R = -11.5 + 0.9\theta_A$) agrees with the data with a correlation coefficient of 0.97. The data thus confirms that there is a linear relationship between the advancing and receding contact angles. It is possible to generalize the relationship between the advancing and receding contact angles for binary and nanofluid droplets since our experimental results are consistent with the range shown in figure \ref{fig:fig6}. For the generalised relationship, if $\theta_A$ is specified, then $\theta_R$ and the maximum Bond number should remain constant. It can be seen that the advancing and receding angles obtained from our experiments for (E 100\% + W 0\%) and (E 80\% + W 20\%) droplets (indicated by the red and green symbols) lie within this range. It is to be noted that our results and earlier experiments have shown that the data is noisy, but the spread of our results falls within the range of those earlier studies. Although the earlier experimental results involved pure fluids on unheated substrates, the current work involves pure ethanol and ethanol-water binary mixture on a heated substrate with nanoparticle loadings. We infer that the correlations given by ElSherbini and Jacobi and linear fit ($\theta_R = -11.5 + 0.9\theta_A$) are reasonably robust for binary droplets and other types of substrate conditions. Despite our limited experimental data, we anticipate that our results will contribute to an eventual generalisation that can explain all experimental results for various droplet types and substrate conditions.

\begin{figure}[h]
\centering
\includegraphics[width=0.7\textwidth]{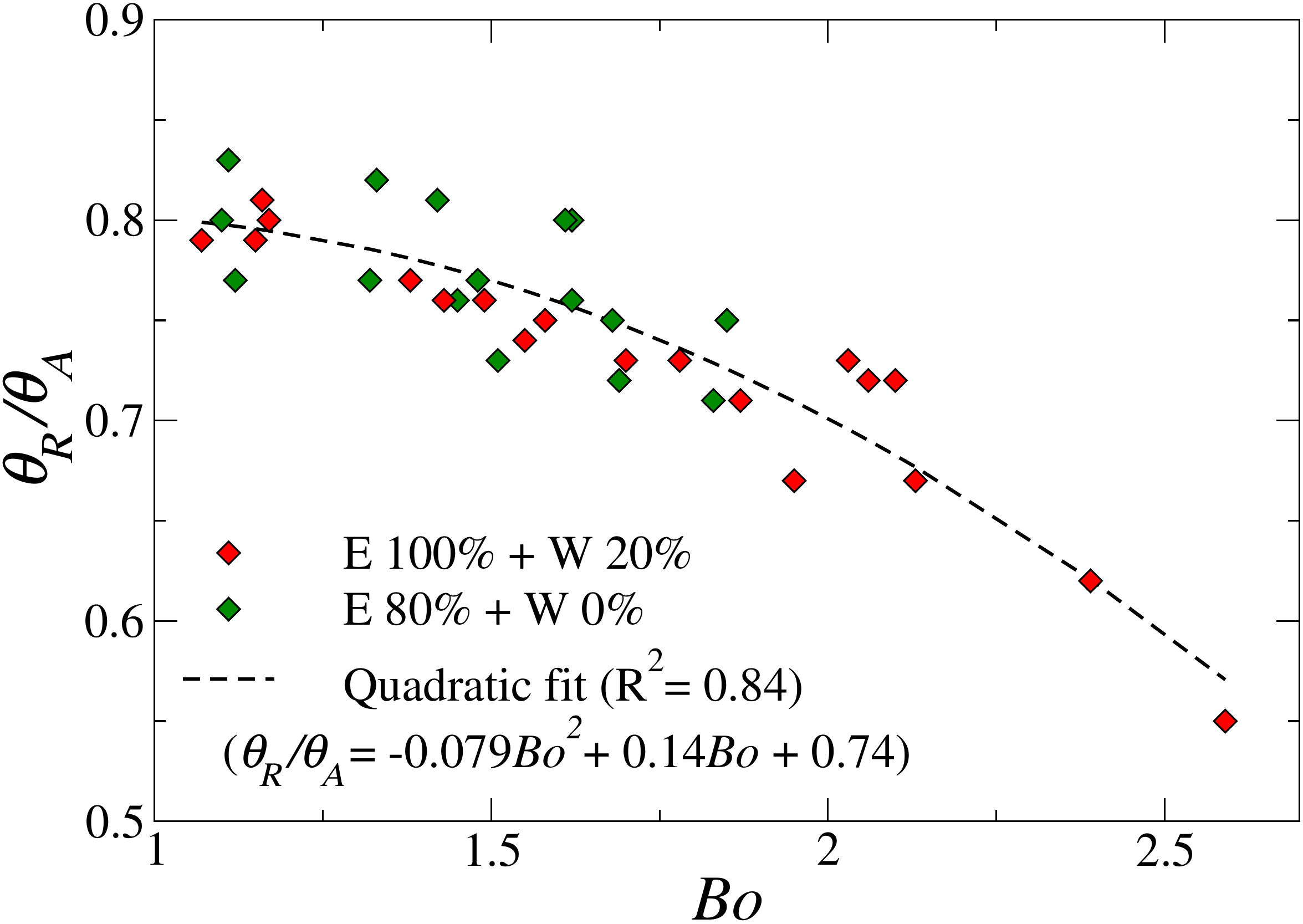}
\caption{Variation of Bond number with a ratio of advancing ($\theta_A$) and receding ($\theta_R$) contact angles. A quadratic curve fits the data with R\textsuperscript{2} = 0.84.}
\label{fig:fig7}
\end{figure}
Figure \ref{fig:fig7} depicts the variation of $\theta_R/\theta_A$ with the Bond number, $Bo$ for (E 100\% + W 0\%) and (E 80\% + W 20\%) droplets with different nanoparticle loadings. A quadratic curve provides a good fit to the data for both (E 100\% + W 0\%) and (E 80\% + W 20\%) droplets for all wt.\%. This quadratic fit with a correlation coefficient of 0.84 is given by
\begin{equation}
    \frac{\theta_R}{\theta_A} = -0.079 Bo^2 + 0.14 Bo + 0.74.
    \label{eq:eq4}
\end{equation}
ElSherbini and Jacobi \cite{elsherbini2006retention} correlated $\theta_{min}$ and $\theta_{max}$ with the Bond number and mentioned that $\theta_{max} \approx \theta_A$. The results indicate that, regardless of the size of the drop, the minimum contact angle in a drop at a critical condition is equal to the receding contact angle $\theta_R$. Thus, in the calculation of the retention force factor, we use $\theta_A$ and $\theta_R$ instead of  $\theta_{max}$ and $\theta_{min}$, respectively.

\subsection{3.3 Retention force factor}

\begin{figure}[h]
\centering
  \includegraphics[width=0.7\textwidth]{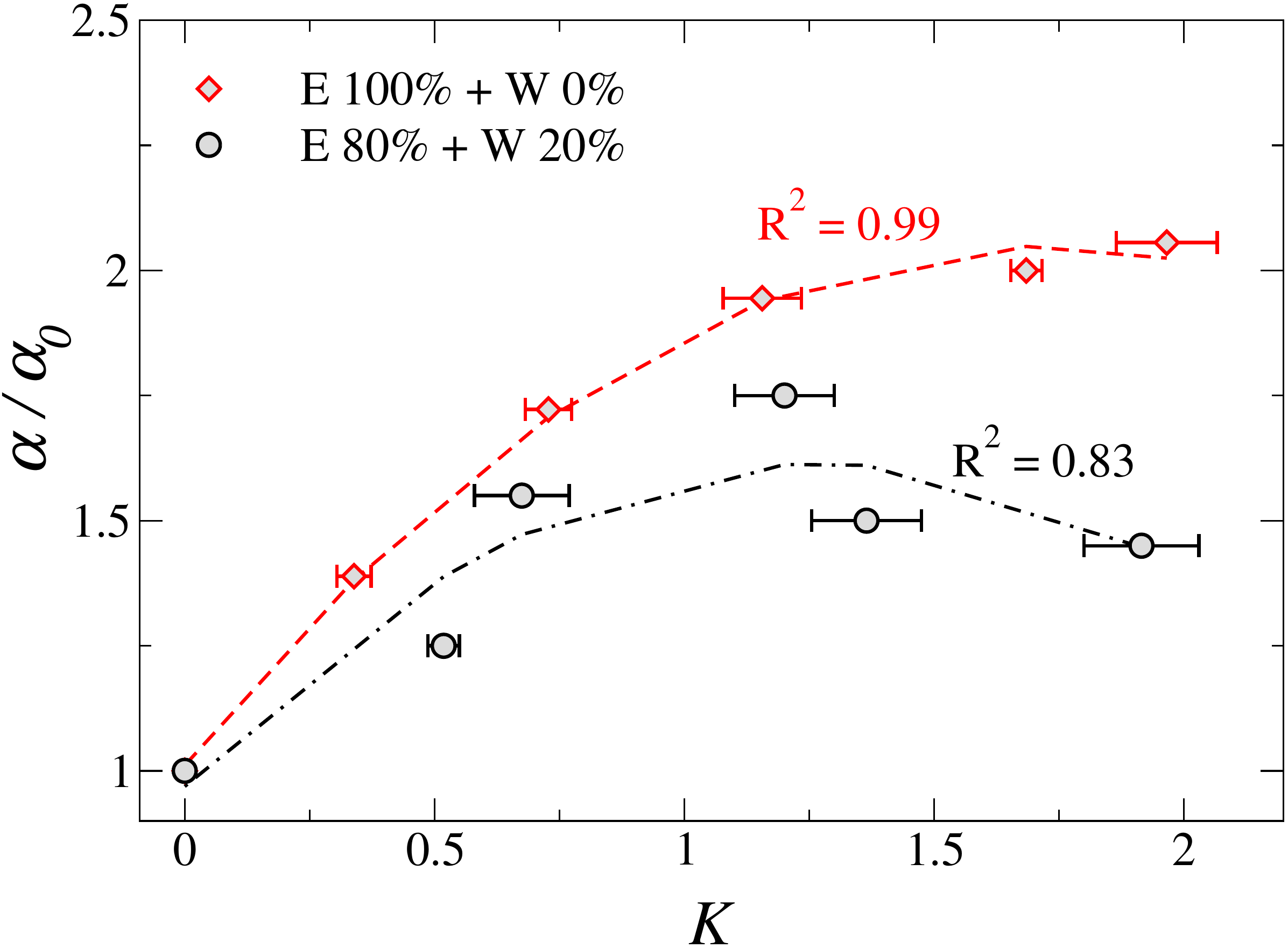}
  \caption{Variation of the modified retention force factor $(K)$ with normalised inclination angle ($\alpha/\alpha_0$) for (E 100\% + W 0\%) and (E 80\% + W 20\%) nanofluid droplets. The data is fitted by quadratic curves with R\textsuperscript{2} values of 0.99 and 0.84 for ethanol and (E 80\% + W 20\%), respectively. The equations of quadratic fits for (E 100\% + W 0\%) and (E 80\% + W 20\%) droplets are ${\alpha}/{\alpha_0} = 1 + 1.2K-0.35K^2$ and ${\alpha}/{\alpha_0} = 1 + 1.2K -0.61K^2$, respectively.}
\label{fig:fig8}
\end{figure}

To study the droplet stability on a critically inclined substrate, the retention force factor is calculated from Eq. (\ref{eq:eq1}) as 
\begin{equation}
    mg \sin \alpha  = k \gamma R (\cos\theta_R - \cos\theta_A),
\label{eq:eq5}
\end{equation}
where $m$ is the mass of the droplet. It is calculated as
\begin{equation}
m = V  \rho_{eff}.
\label{eq:eq6}
\end{equation}
Here, $V$ is volume of the droplet. By assuming spherical cap assumption, the volume of an asymmetrical droplet can be calculated using \cite{elsherbini2006retention},
\begin{equation}
V  = {\pi R^3 \over 3} { (2 -3 \cos \theta_{\rm avg} + \cos^3 \theta_{\rm avg}) \over\sin^3\theta_{\rm avg}},
\label{eq:eq7}
\end{equation}
where $\theta_{\rm avg} = (\theta_A + \theta_R$)/2. Figure \ref{fig:fig8} depicts the retention force factor ($k$) for binary nanofluid droplets obtained from Eq. (\ref{eq:eq5}) in terms of $K$. The modified retention force factor, $K$, is given by $k\times (V_m/V)\times C$, where $V_m$ is the mean volume of droplets considering all the cases and $C$ is the nanoparticle concentration in wt.\%. It is to be noted that the initial volume of the droplets for different conditions varies slightly due to experiment uncertainty. Thus, we multiply $k$ by the volume correction factor $V_m/V$. The modified retention force factor, $K$, is plotted against the normalized critical inclination angle (i.e. the critical angle normalized with the critical angle of no loading case). The results for (E 100\% + W 0\%) and (E 80\% + W 20\%) droplets are fitted by quadratic fit with a correlation coefficient of 0.99 and 0.83, respectively. The equations for quadratic fits are given by
\\
For (E 100\% + W 0\%) droplets:
\begin{equation}
{\alpha} /{\alpha_0} = 1 + 1.2K -0.35K^2,  
\label{eq:eq8}
\end{equation}
and for (E 80\% + W 20\%) droplets:
\begin{equation}
{\alpha} /{\alpha_0} = 1 + 1.2K -0.61K^2. 
\label{eq:eq9}
\end{equation}

The constants and linear terms in Eqs. (\ref{eq:eq8}) and (\ref{eq:eq9}) are the same for both (E 100\% + W 0\%) and (E 80\% + W 20\%) droplets. For droplets with no nanoparticle loading, $C$ is zero, and consequently, $K$ becomes zero from its definition mentioned above. The proposed correlation focuses on the change in a droplet's critical angle driven by the addition of nanoparticles in comparison to the no-loading condition. The linear term represents the increase in the critical angle due to an increase in friction and surface tension caused by the addition of the nanoparticles. The expressions show that for lower loading wt.\%, the effect of adding nanoparticles is the same for both compositions since the linear term has the same constant for both droplets. On the other hand, the quadratic term is negative and decreases the critical angle with increasing nanoparticle loading. This term comes in due to the increase in weight of the nanofluid due to the addition of nanoparticles that increases the gravitational force component. Thus this term becomes important only at higher wt.\% of nanoparticle loading. The constant associated with this quadratic term is also different for pure and binary fluids. This is likely because the balance between surface tension and gravitational forces differs for critically inclined droplets of different compositions. The coefficient of $K^2$ for pure fluid (E 100\% + W 0\%) ($a_1$) and binary fluid (E 80\% + W 20\%) ($a_2$) are related as $a_2 = 1.74 a_1$. The $K$ values are influenced by the inclination, composition of the mixture, and concentration of the nanoparticles added, as shown in figure \ref{fig:fig8}.  The value of $K$ for (E 100\% + W 0\%) droplet increases with increasing inclination angle and nanoparticle concentration. For (E 80\% + W 20\%) droplet, the value of $K$ increases initially up to 0.6 wt.\% as the critical angle increases. After 0.6 wt.\%, a slight decrease in the critical angle is observed despite this value of $K$ increases as other terms in Eq. \ref{eq:eq5} like $(\cos \theta_R - \cos \theta_A)$ and $m$ dominate.

\subsection{3.4 Evaporation dynamics: thermal profiles}

\begin{figure}[h]
\centering
\includegraphics[width=0.9\textwidth]{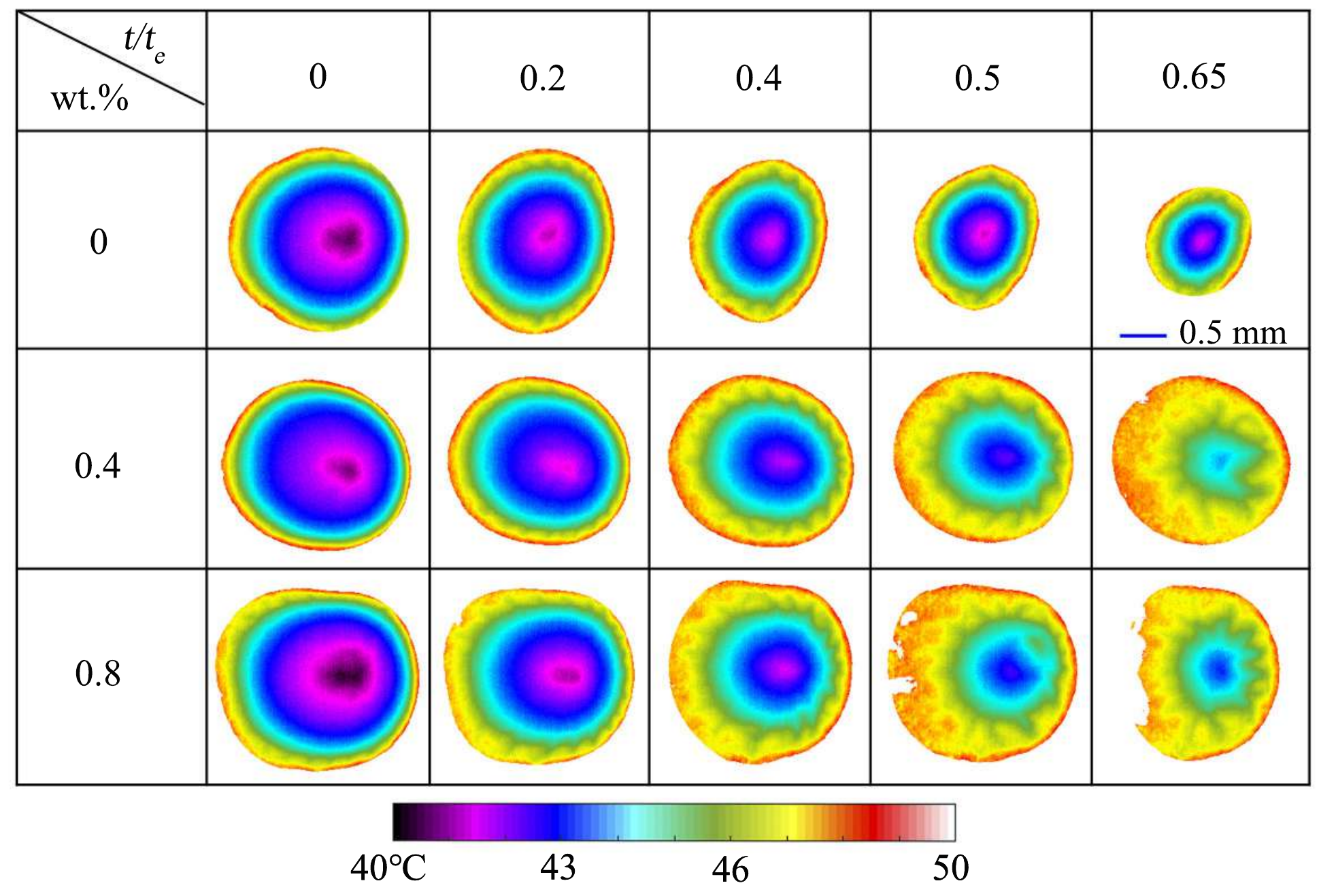}
\caption{Temporal evolution of the temperature contours of an ethanol (E 100\% + W 0\%) droplet for different nanoparticle loadings (wt.\%). The color bar shows the temperature variation.}
\label{fig:fig9}
\end{figure}

In this section, we discuss the thermal patterns of the evaporating sessile droplets of (E 100\% + W 0\%) and (E 80\% + W 20\%) compositions laden with different nanoparticle loadings. The substrate is maintained at $T_s = 50^\circ$ and inclined at the respective critical angle of different droplets. Figure \ref{fig:fig9} shows the temporal variations of the temperature contours of an (E 100\% + W 0\%) droplet with no loading, 0.4 wt.\%, and 0.8 wt.\% loadings. Here, $t/t_e$ is the normalized time, wherein $t_e$ is the lifetime of the droplet and $t=0$ represents the instant when the droplet is placed on the substrate. It can be observed that the wetting diameter of the (E 100\% + W 0\%) droplet without nanoparticle loading shrinks (i.e the wetting diameter decreases with time) whereas the droplet with loading shows a pinned behavior for the majority of its lifetime. In the loading condition, we observe more hydrothermal waves and instabilities since the droplet is pinning and the addition of nanoparticles accelerates the evaporation process. This demonstrates that in droplets laden with nanoparticles, the thermocapillary forces due to the surface tension gradients produce more vigorous Marangoni convection. The differences in the surface tension brought on by the temperature gradients cause the interfacial waves to move from warmer (lower surface tension) to colder (higher surface tension) zones. As a result of these actions, the droplet tends to spread along the direction of inclination. We observe more elongation of the droplet in the case of 0.8 wt.\% than 0.4 wt.\% case due to its higher critical angle of inclination. This resulted in an earlier breakdown of the droplet at the receding side as more fluid volume shifted to the advancing side.

\begin{figure}[h]
\centering
\includegraphics[width=0.9\textwidth]{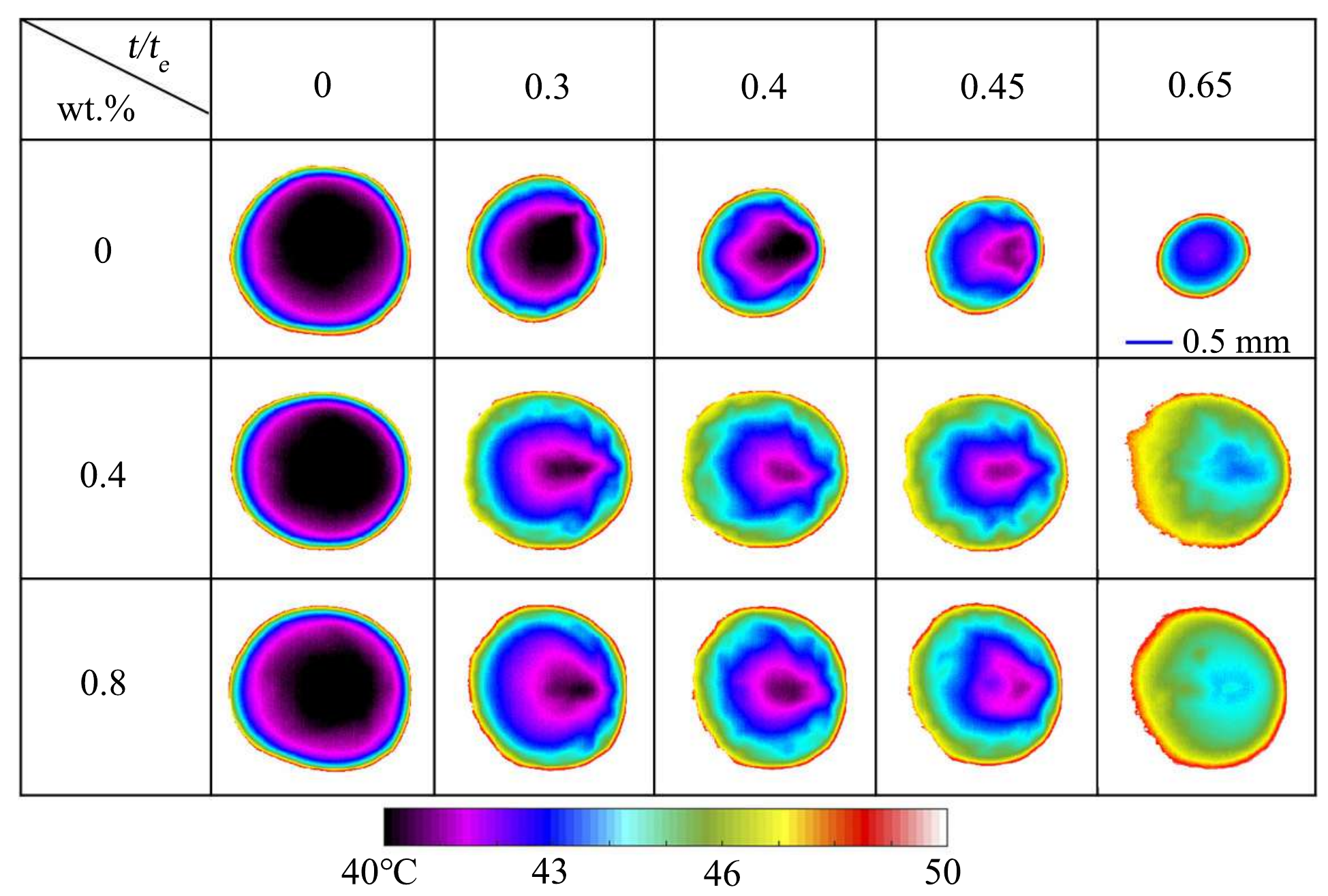}
\caption{Temporal evolution of the temperature contours of a (E 80\% + W 20\%) droplet for different nanoparticle loadings (wt.\%). The color bar shows the temperature variation.}
\label{fig:fig10}
\end{figure}

Figure \ref{fig:fig10} depicts the temporal evolution of the temperature contours of an (E 80\% + W 20\%) droplet for different nanoparticle loadings (wt.\%). It can be seen that, similar to the pure ethanol droplet shown in figure \ref{fig:fig9}, the (E 80\% + W 20\%) droplet with nanoparticles loadings also shows a pinned nature whereas no loading droplet shrinks as its wetting diameter decreases with time (figure \ref{fig:fig10}). For this composition, due to the small difference in the critical angle values between 0.4 wt.\% and 0.8 wt.\%, no significant variation is observed in the thermal patterns. The hydrothermal waves and instability are less intense in an (E 80\% + W 20\%) droplet than an (E 100\% + W 0\%) as the gradient of surface tension $\frac{d\sigma}{dT}$ for water is lower than that of ethanol \cite{vazquez1995surface}, that is, $\left(\frac{d\sigma}{dT}\right)_{water} < \left(\frac{d\sigma}{dT}\right)_{ethanol}$.  

\section{4. Conclusions}
\label{sec:conclusion}

The droplet stability in terms of the retention force factor at their respective critical angles is studied for pure ethanol (E 100\% + W 0\%) and binary (E 80\% + W 20\%) sessile droplets laden with alumina (Al$_2$O$_3$) nanoparticles. Shadowgraphy and infrared imaging techniques are employed to investigate the dynamics, and experiments are conducted in a customized goniometer. The images are post-processed using \textsc{Matlab}$^{\circledR}$ and U-net architecture based on a convolution neural network (machine learning technique). It is observed that the critical angle of inclination increases with an increase in nanoparticle loading (wt.\%) for (E 100\% + W 0\%) droplets. However, for (E 80\% + W 20\%) droplets, the critical angle decreases slightly after 0.6 wt.\% as the gravitational force dominates the surface tension force. With the increase in the critical angle of inclination, the Bond number also increases. The values of advancing and receding angles from different sources show that the relationship between the advancing and receding contact angles can be generalized for a given range. Our experimental advancing and receding contact angle values fall under this range. A quadratic fit with R\textsuperscript{2} value of 0.84 fits the relationship between the ratio of the advancing to receding contact angles and the Bond number for both (E 100\% + W 0\%) and (E 80\% + W 20\%) droplets. The retention force factor is calculated for different nanoparticle loadings for pure and binary fluid droplets. The droplet composition, nanoparticles loading, and the critical angle of inclination affect the retention force factor. For (E 100\% + W 0\%) droplets, the retention force factor increases with an increase in nanoparticle loading and the critical inclination angle. The data are fitted by a quadratic fit and the correlation for retention force factor and the critical angle between (E 100\% + W 0\%) and (E 80\% + W 20\%) droplets is found. Infrared images of the evaporating droplets are presented for different loading conditions. The droplets with nanoparticle loading show richer hydrothermal waves than the droplets without loading, and these waves are more intense in droplets with higher ethanol concentration. The droplets elongated more towards the receding side due to body force, resulting in an earlier breakdown of the droplet at the receding side.\\

%     \section*{Supporting Information}
%     
%    \begin{itemize} 
%     \item Represention of the advancing ($\theta_A$) and receding contact angles ($\theta_R$) of a droplet on an inclined plate.
%      \item The values of advancing ($\theta_A$) and receding ($\theta_R$) contact angles in degree for (E100\% + W0\%) and (E80\% + W20\%) droplets placed their respective critical angles for different values of nanoparticle loading (wt.\%).
%      \item The values of the critical angle of inclination ($\alpha$, in degree) for (E100\% + W0\%) and (E80\% + W20\%) droplets at the onset of sliding for different values of nanoparticle loading (wt.\%).
%    \end{itemize} 

\noindent{\bf Acknowledgement:} K.C.S. thanks Science \& Engineering Research Board, India, for financial support through grant number: CRG/2020/000507. We also thank Dr Lakshmana Dora Chandrala for his help in employing the machine learning method for post-processing.

%\bibliography{bibl}

\begin{mcitethebibliography}{58}
\providecommand*\natexlab[1]{#1}
\providecommand*\mciteSetBstSublistMode[1]{}
\providecommand*\mciteSetBstMaxWidthForm[2]{}
\providecommand*\mciteBstWouldAddEndPuncttrue
  {\def\EndOfBibitem{\unskip.}}
\providecommand*\mciteBstWouldAddEndPunctfalse
  {\let\EndOfBibitem\relax}
\providecommand*\mciteSetBstMidEndSepPunct[3]{}
\providecommand*\mciteSetBstSublistLabelBeginEnd[3]{}
\providecommand*\EndOfBibitem{}
\mciteSetBstSublistMode{f}
\mciteSetBstMaxWidthForm{subitem}{(\alph{mcitesubitemcount})}
\mciteSetBstSublistLabelBeginEnd
  {\mcitemaxwidthsubitemform\space}
  {\relax}
  {\relax}

\bibitem[Lim \latin{et~al.}(2012)Lim, Yang, Lee, Chung, and
  Hong]{lim2012deposit}
Lim,~T.; Yang,~J.; Lee,~S.; Chung,~J.; Hong,~D. Deposit pattern of inkjet
  printed pico-liter droplet. \emph{Int. J. Pr. Eng. Man.} \textbf{2012},
  \emph{13}, 827--833\relax
\mciteBstWouldAddEndPuncttrue
\mciteSetBstMidEndSepPunct{\mcitedefaultmidpunct}
{\mcitedefaultendpunct}{\mcitedefaultseppunct}\relax
\EndOfBibitem
\bibitem[Kim \latin{et~al.}(2006)Kim, Jeong, Park, and Moon]{kim2006direct}
Kim,~D.; Jeong,~S.; Park,~B.~K.; Moon,~J. Direct writing of silver conductive
  patterns: Improvement of film morphology and conductance by controlling
  solvent compositions. \emph{Appl. Phys. Lett.} \textbf{2006}, \emph{89},
  264101\relax
\mciteBstWouldAddEndPuncttrue
\mciteSetBstMidEndSepPunct{\mcitedefaultmidpunct}
{\mcitedefaultendpunct}{\mcitedefaultseppunct}\relax
\EndOfBibitem
\bibitem[Park and Moon(2006)Park, and Moon]{park2006control}
Park,~J.; Moon,~J. Control of colloidal particle deposit patterns within
  picoliter droplets ejected by ink-jet printing. \emph{Langmuir}
  \textbf{2006}, \emph{22}, 3506--3513\relax
\mciteBstWouldAddEndPuncttrue
\mciteSetBstMidEndSepPunct{\mcitedefaultmidpunct}
{\mcitedefaultendpunct}{\mcitedefaultseppunct}\relax
\EndOfBibitem
\bibitem[Lim \latin{et~al.}(2009)Lim, Han, Chung, Chung, Ko, and
  Grigoropoulos]{lim2009experimental}
Lim,~T.; Han,~S.; Chung,~J.; Chung,~J.~T.; Ko,~S.; Grigoropoulos,~C.~P.
  Experimental study on spreading and evaporation of inkjet printed pico-liter
  droplet on a heated substrate. \emph{Int. J. Heat Mass Transf.}
  \textbf{2009}, \emph{52}, 431--441\relax
\mciteBstWouldAddEndPuncttrue
\mciteSetBstMidEndSepPunct{\mcitedefaultmidpunct}
{\mcitedefaultendpunct}{\mcitedefaultseppunct}\relax
\EndOfBibitem
\bibitem[Koo \latin{et~al.}(2006)Koo, Chen, Pan, Chou, Wu, Chang, and
  Kawai]{koo2006fabrication}
Koo,~H.; Chen,~M.; Pan,~P.; Chou,~L.; Wu,~F.; Chang,~S.; Kawai,~T. Fabrication
  and chromatic characteristics of the greenish {LCD} colour-filter layer with
  nano-particle ink using inkjet printing technique. \emph{Displays}
  \textbf{2006}, \emph{27}, 124--129\relax
\mciteBstWouldAddEndPuncttrue
\mciteSetBstMidEndSepPunct{\mcitedefaultmidpunct}
{\mcitedefaultendpunct}{\mcitedefaultseppunct}\relax
\EndOfBibitem
\bibitem[Tekin \latin{et~al.}(2004)Tekin, de~Gans, and Schubert]{tekin2004ink}
Tekin,~E.; de~Gans,~B.~J.; Schubert,~U.~S. Ink-jet printing of polymers--from
  single dots to thin film libraries. \emph{J. Mater. Chem.} \textbf{2004},
  \emph{14}, 2627--2632\relax
\mciteBstWouldAddEndPuncttrue
\mciteSetBstMidEndSepPunct{\mcitedefaultmidpunct}
{\mcitedefaultendpunct}{\mcitedefaultseppunct}\relax
\EndOfBibitem
\bibitem[de~Gans and Schubert(2004)de~Gans, and Schubert]{de2004inkjet}
de~Gans,~B.~J.; Schubert,~U.~S. Inkjet printing of well-defined polymer dots
  and arrays. \emph{Langmuir} \textbf{2004}, \emph{20}, 7789--7793\relax
\mciteBstWouldAddEndPuncttrue
\mciteSetBstMidEndSepPunct{\mcitedefaultmidpunct}
{\mcitedefaultendpunct}{\mcitedefaultseppunct}\relax
\EndOfBibitem
\bibitem[Yanagisawa \latin{et~al.}(2014)Yanagisawa, Sakai, Isobe, Matsushita,
  and Nakajima]{yanagisawa2014investigation}
Yanagisawa,~K.; Sakai,~M.; Isobe,~T.; Matsushita,~S.; Nakajima,~A.
  Investigation of droplet jumping on superhydrophobic coatings during dew
  condensation by the observation from two directions. \emph{Appl. Surf. Sci.}
  \textbf{2014}, \emph{315}, 212--221\relax
\mciteBstWouldAddEndPuncttrue
\mciteSetBstMidEndSepPunct{\mcitedefaultmidpunct}
{\mcitedefaultendpunct}{\mcitedefaultseppunct}\relax
\EndOfBibitem
\bibitem[Deegan \latin{et~al.}(1997)Deegan, Bakajin, Dupont, Huber, Nagel, and
  Witten]{deegan1997capillary}
Deegan,~R.~D.; Bakajin,~O.; Dupont,~T.~F.; Huber,~G.; Nagel,~S.~R.;
  Witten,~T.~A. Capillary flow as the cause of ring stains from dried liquid
  drops. \emph{Nature} \textbf{1997}, \emph{389}, 827--829\relax
\mciteBstWouldAddEndPuncttrue
\mciteSetBstMidEndSepPunct{\mcitedefaultmidpunct}
{\mcitedefaultendpunct}{\mcitedefaultseppunct}\relax
\EndOfBibitem
\bibitem[Dugas \latin{et~al.}(2005)Dugas, Broutin, and
  Souteyrand]{dugas2005droplet}
Dugas,~V.; Broutin,~J.; Souteyrand,~E. Droplet evaporation study applied to DNA
  chip manufacturing. \emph{Langmuir} \textbf{2005}, \emph{21},
  9130--9136\relax
\mciteBstWouldAddEndPuncttrue
\mciteSetBstMidEndSepPunct{\mcitedefaultmidpunct}
{\mcitedefaultendpunct}{\mcitedefaultseppunct}\relax
\EndOfBibitem
\bibitem[Lee \latin{et~al.}(2006)Lee, Cho, Huh, Ko, Lee, Jang, Lee, Kang, and
  Choi]{lee2006electrohydrodynamic}
Lee,~J.-G.; Cho,~H.-J.; Huh,~N.; Ko,~C.; Lee,~W.-C.; Jang,~Y.-H.; Lee,~B.~S.;
  Kang,~I.~S.; Choi,~J.-W. Electrohydrodynamic (EHD) dispensing of nanoliter
  DNA droplets for microarrays. \emph{Biosens. Bioelectron.} \textbf{2006},
  \emph{21}, 2240--2247\relax
\mciteBstWouldAddEndPuncttrue
\mciteSetBstMidEndSepPunct{\mcitedefaultmidpunct}
{\mcitedefaultendpunct}{\mcitedefaultseppunct}\relax
\EndOfBibitem
\bibitem[Balusamy \latin{et~al.}(2021)Balusamy, Banerjee, and
  Sahu]{balusamy2021lifetime}
Balusamy,~S.; Banerjee,~S.; Sahu,~K.~C. Lifetime of sessile saliva droplets in
  the context of SARS-CoV-2. \emph{Int. Commun. Heat Mass Transf.}
  \textbf{2021}, \emph{123}, 105178\relax
\mciteBstWouldAddEndPuncttrue
\mciteSetBstMidEndSepPunct{\mcitedefaultmidpunct}
{\mcitedefaultendpunct}{\mcitedefaultseppunct}\relax
\EndOfBibitem
\bibitem[Brutin \latin{et~al.}(2012)Brutin, Sobac, and
  Nicloux]{brutin2012influence}
Brutin,~D.; Sobac,~B.; Nicloux,~C. Influence of substrate nature on the
  evaporation of a sessile drop of blood. \emph{J. Heat Transf.} \textbf{2012},
  \emph{134}\relax
\mciteBstWouldAddEndPuncttrue
\mciteSetBstMidEndSepPunct{\mcitedefaultmidpunct}
{\mcitedefaultendpunct}{\mcitedefaultseppunct}\relax
\EndOfBibitem
\bibitem[Zeid \latin{et~al.}(2013)Zeid, Vicente, and Brutin]{zeid2013influence}
Zeid,~W.~B.; Vicente,~J.; Brutin,~D. Influence of evaporation rate on cracks’
  formation of a drying drop of whole blood. \emph{Colloids Surf. A:
  Physicochem. Eng. Asp.} \textbf{2013}, \emph{432}, 139--146\relax
\mciteBstWouldAddEndPuncttrue
\mciteSetBstMidEndSepPunct{\mcitedefaultmidpunct}
{\mcitedefaultendpunct}{\mcitedefaultseppunct}\relax
\EndOfBibitem
\bibitem[Lanotte \latin{et~al.}(2017)Lanotte, Laux, Charlot, and
  Abkarian]{lanotte2017role}
Lanotte,~L.; Laux,~D.; Charlot,~B.; Abkarian,~M. Role of red cells and plasma
  composition on blood sessile droplet evaporation. \emph{Phys. Rev. E}
  \textbf{2017}, \emph{96}, 053114\relax
\mciteBstWouldAddEndPuncttrue
\mciteSetBstMidEndSepPunct{\mcitedefaultmidpunct}
{\mcitedefaultendpunct}{\mcitedefaultseppunct}\relax
\EndOfBibitem
\bibitem[Sefiane(2014)]{sefiane2014patterns}
Sefiane,~K. Patterns from drying drops. \emph{Adv. Colloid Interface Sci.}
  \textbf{2014}, \emph{206}, 372--381\relax
\mciteBstWouldAddEndPuncttrue
\mciteSetBstMidEndSepPunct{\mcitedefaultmidpunct}
{\mcitedefaultendpunct}{\mcitedefaultseppunct}\relax
\EndOfBibitem
\bibitem[Erbil(2015)]{erbil2015control}
Erbil,~H.~Y. Control of stain geometry by drop evaporation of surfactant
  containing dispersions. \emph{Adv. Colloid Interface Sci.} \textbf{2015},
  \emph{222}, 275--290\relax
\mciteBstWouldAddEndPuncttrue
\mciteSetBstMidEndSepPunct{\mcitedefaultmidpunct}
{\mcitedefaultendpunct}{\mcitedefaultseppunct}\relax
\EndOfBibitem
\bibitem[Zhong and Duan(2014)Zhong, and Duan]{zhong2014evaporation}
Zhong,~X.; Duan,~F. Evaporation of sessile droplets affected by graphite
  nanoparticles and binary base fluids. \emph{J. Phys. Chem. B.} \textbf{2014},
  \emph{118}, 13636--13645\relax
\mciteBstWouldAddEndPuncttrue
\mciteSetBstMidEndSepPunct{\mcitedefaultmidpunct}
{\mcitedefaultendpunct}{\mcitedefaultseppunct}\relax
\EndOfBibitem
\bibitem[Zhong and Duan(2016)Zhong, and Duan]{zhong2016flow}
Zhong,~X.; Duan,~F. Flow regime and deposition pattern of evaporating binary
  mixture droplet suspended with particles. \emph{Eur. Phys. J. E Soft Matter}
  \textbf{2016}, \emph{39}, 18\relax
\mciteBstWouldAddEndPuncttrue
\mciteSetBstMidEndSepPunct{\mcitedefaultmidpunct}
{\mcitedefaultendpunct}{\mcitedefaultseppunct}\relax
\EndOfBibitem
\bibitem[Parsa \latin{et~al.}(2017)Parsa, Boubaker, Harmand, Sefiane,
  Bigerelle, and Deltombe]{parsa2017patterns}
Parsa,~M.; Boubaker,~R.; Harmand,~S.; Sefiane,~K.; Bigerelle,~M.; Deltombe,~R.
  Patterns from dried water-butanol binary-based nanofluid drops. \emph{J.
  Nanopart. Res.} \textbf{2017}, \emph{19}, 268\relax
\mciteBstWouldAddEndPuncttrue
\mciteSetBstMidEndSepPunct{\mcitedefaultmidpunct}
{\mcitedefaultendpunct}{\mcitedefaultseppunct}\relax
\EndOfBibitem
\bibitem[Hari~Govindha \latin{et~al.}(2022)Hari~Govindha, Katre, Balusamy,
  Banerjee, and Sahu]{hari2022counter}
Hari~Govindha,~A.; Katre,~P.; Balusamy,~S.; Banerjee,~S.; Sahu,~K.~C.
  Counter-Intuitive Evaporation in Nanofluids Droplets due to Stick-Slip
  Nature. \emph{Langmuir} \textbf{2022}, \relax
\mciteBstWouldAddEndPunctfalse
\mciteSetBstMidEndSepPunct{\mcitedefaultmidpunct}
{}{\mcitedefaultseppunct}\relax
\EndOfBibitem
\bibitem[Li \latin{et~al.}(2018)Li, Ji, Sun, Lan, and Wang]{li2018pattern}
Li,~W.; Ji,~W.; Sun,~H.; Lan,~D.; Wang,~Y. Pattern formation in drying sessile
  and pendant droplet: Interactions of gravity settling, interface shrinkage,
  and capillary flow. \emph{Langmuir} \textbf{2018}, \emph{35}, 113--119\relax
\mciteBstWouldAddEndPuncttrue
\mciteSetBstMidEndSepPunct{\mcitedefaultmidpunct}
{\mcitedefaultendpunct}{\mcitedefaultseppunct}\relax
\EndOfBibitem
\bibitem[Mondal \latin{et~al.}(2018)Mondal, Semwal, Kumar, Thampi, and
  Basavaraj]{mondal2018patterns}
Mondal,~R.; Semwal,~S.; Kumar,~P.~L.; Thampi,~S.~P.; Basavaraj,~M.~G. Patterns
  in drying drops dictated by curvature-driven particle transport.
  \emph{Langmuir} \textbf{2018}, \emph{34}, 11473--11483\relax
\mciteBstWouldAddEndPuncttrue
\mciteSetBstMidEndSepPunct{\mcitedefaultmidpunct}
{\mcitedefaultendpunct}{\mcitedefaultseppunct}\relax
\EndOfBibitem
\bibitem[Gopu \latin{et~al.}(2020)Gopu, Rathod, Namangalam, Pujala, Kumar, and
  Mampallil]{gopu2020evaporation}
Gopu,~M.; Rathod,~S.; Namangalam,~U.; Pujala,~R.~K.; Kumar,~S.~S.;
  Mampallil,~D. Evaporation of inclined drops: Formation of asymmetric ring
  patterns. \emph{Langmuir} \textbf{2020}, \emph{36}, 8137--8143\relax
\mciteBstWouldAddEndPuncttrue
\mciteSetBstMidEndSepPunct{\mcitedefaultmidpunct}
{\mcitedefaultendpunct}{\mcitedefaultseppunct}\relax
\EndOfBibitem
\bibitem[ElSherbini and Jacobi(2006)ElSherbini, and
  Jacobi]{elsherbini2006retention}
ElSherbini,~A.; Jacobi,~A. Retention forces and contact angles for critical
  liquid drops on non-horizontal surfaces. \emph{J. Colloid Interf. Sci.}
  \textbf{2006}, \emph{299}, 841--849\relax
\mciteBstWouldAddEndPuncttrue
\mciteSetBstMidEndSepPunct{\mcitedefaultmidpunct}
{\mcitedefaultendpunct}{\mcitedefaultseppunct}\relax
\EndOfBibitem
\bibitem[Extrand and Kumagai(1995)Extrand, and Kumagai]{extrand1995liquid}
Extrand,~C.~W.; Kumagai,~Y. Liquid drops on an inclined plane: the relation
  between contact angles, drop shape, and retentive force. \emph{J. Colloid
  Interf. Sci.} \textbf{1995}, \emph{170}, 515--521\relax
\mciteBstWouldAddEndPuncttrue
\mciteSetBstMidEndSepPunct{\mcitedefaultmidpunct}
{\mcitedefaultendpunct}{\mcitedefaultseppunct}\relax
\EndOfBibitem
\bibitem[Janardan and Panchagnula(2014)Janardan, and
  Panchagnula]{janardan2014effect}
Janardan,~N.; Panchagnula,~M.~V. Effect of the initial conditions on the onset
  of motion in sessile drops on tilted plates. \emph{Colloids Surf. A
  Physicochem. Eng. Asp.} \textbf{2014}, \emph{456}, 238--245\relax
\mciteBstWouldAddEndPuncttrue
\mciteSetBstMidEndSepPunct{\mcitedefaultmidpunct}
{\mcitedefaultendpunct}{\mcitedefaultseppunct}\relax
\EndOfBibitem
\bibitem[Ding \latin{et~al.}(2020)Ding, Jia, Peng, and Guo]{ding2020critical}
Ding,~Y.; Jia,~L.; Peng,~Q.; Guo,~J. Critical sliding angle of water droplet on
  parallel hydrophobic grooved surface. \emph{Colloids Surf. A: Physicochem.
  Eng. Asp.} \textbf{2020}, \emph{585}, 124083\relax
\mciteBstWouldAddEndPuncttrue
\mciteSetBstMidEndSepPunct{\mcitedefaultmidpunct}
{\mcitedefaultendpunct}{\mcitedefaultseppunct}\relax
\EndOfBibitem
\bibitem[Yilbas \latin{et~al.}(2017)Yilbas, Al-Sharafi, Ali, and
  Al-Aqeeli]{yilbas2017dynamics}
Yilbas,~B.~S.; Al-Sharafi,~A.; Ali,~H.; Al-Aqeeli,~N. Dynamics of a water
  droplet on a hydrophobic inclined surface: influence of droplet size and
  surface inclination angle on droplet rolling. \emph{RSC Adv.} \textbf{2017},
  \emph{7}, 48806--48818\relax
\mciteBstWouldAddEndPuncttrue
\mciteSetBstMidEndSepPunct{\mcitedefaultmidpunct}
{\mcitedefaultendpunct}{\mcitedefaultseppunct}\relax
\EndOfBibitem
\bibitem[Kim \latin{et~al.}(2002)Kim, Lee, and Kang]{kim2002sliding}
Kim,~H.; Lee,~H.~J.; Kang,~B.~H. Sliding of liquid drops down an inclined solid
  surface. \emph{J. Colloid Interf. Sci.} \textbf{2002}, \emph{247},
  372--380\relax
\mciteBstWouldAddEndPuncttrue
\mciteSetBstMidEndSepPunct{\mcitedefaultmidpunct}
{\mcitedefaultendpunct}{\mcitedefaultseppunct}\relax
\EndOfBibitem
\bibitem[Annapragada \latin{et~al.}(2012)Annapragada, Murthy, and
  Garimella]{annapragada2012droplet}
Annapragada,~S.~R.; Murthy,~J.~Y.; Garimella,~S.~V. Droplet retention on an
  incline. \emph{Int. J. Heat Mass Transf.} \textbf{2012}, \emph{55},
  1457--1465\relax
\mciteBstWouldAddEndPuncttrue
\mciteSetBstMidEndSepPunct{\mcitedefaultmidpunct}
{\mcitedefaultendpunct}{\mcitedefaultseppunct}\relax
\EndOfBibitem
\bibitem[Chou \latin{et~al.}(2012)Chou, Hong, Sheng, and Tsao]{chou2012drops}
Chou,~T.-H.; Hong,~S.-J.; Sheng,~Y.-J.; Tsao,~H.-K. Drops sitting on a tilted
  plate: receding and advancing pinning. \emph{Langmuir} \textbf{2012},
  \emph{28}, 5158--5166\relax
\mciteBstWouldAddEndPuncttrue
\mciteSetBstMidEndSepPunct{\mcitedefaultmidpunct}
{\mcitedefaultendpunct}{\mcitedefaultseppunct}\relax
\EndOfBibitem
\bibitem[R{\'\i}os-L{\'o}pez \latin{et~al.}(2018)R{\'\i}os-L{\'o}pez,
  Evgenidis, Kostoglou, Zabulis, and Karapantsios]{rios2018effect}
R{\'\i}os-L{\'o}pez,~I.; Evgenidis,~S.; Kostoglou,~M.; Zabulis,~X.;
  Karapantsios,~T.~D. Effect of initial droplet shape on the tangential force
  required for spreading and sliding along a solid surface. \emph{Colloids
  Surf. A: Physicochem. Eng. Asp.} \textbf{2018}, \emph{549}, 164--173\relax
\mciteBstWouldAddEndPuncttrue
\mciteSetBstMidEndSepPunct{\mcitedefaultmidpunct}
{\mcitedefaultendpunct}{\mcitedefaultseppunct}\relax
\EndOfBibitem
\bibitem[Furmidge(1962)]{furmidge1962studies}
Furmidge,~C. Studies at phase interfaces. I. The sliding of liquid drops on
  solid surfaces and a theory for spray retention. \emph{J. Colloid Interf.
  Sci.} \textbf{1962}, \emph{17}, 309--324\relax
\mciteBstWouldAddEndPuncttrue
\mciteSetBstMidEndSepPunct{\mcitedefaultmidpunct}
{\mcitedefaultendpunct}{\mcitedefaultseppunct}\relax
\EndOfBibitem
\bibitem[Ozturk and Erbil(2018)Ozturk, and Erbil]{ozturk2018evaporation}
Ozturk,~T.; Erbil,~H.~Y. Evaporation of water-ethanol binary sessile drop on
  fluoropolymer surfaces: Influence of relative humidity. \emph{Colloids Surf.
  A} \textbf{2018}, \emph{553}, 327--336\relax
\mciteBstWouldAddEndPuncttrue
\mciteSetBstMidEndSepPunct{\mcitedefaultmidpunct}
{\mcitedefaultendpunct}{\mcitedefaultseppunct}\relax
\EndOfBibitem
\bibitem[Ozturk and Erbil(2020)Ozturk, and Erbil]{ozturk2020simple}
Ozturk,~T.; Erbil,~H.~Y. Simple Model for Diffusion-Limited Drop Evaporation of
  Binary Liquids from Physical Properties of the Components: Ethanol--Water
  Example. \emph{Langmuir} \textbf{2020}, \emph{36}, 1357--1371\relax
\mciteBstWouldAddEndPuncttrue
\mciteSetBstMidEndSepPunct{\mcitedefaultmidpunct}
{\mcitedefaultendpunct}{\mcitedefaultseppunct}\relax
\EndOfBibitem
\bibitem[Li \latin{et~al.}(2019)Li, Diddens, Lv, Wijshoff, Versluis, and
  Lohse]{li2019gravitational}
Li,~Y.; Diddens,~C.; Lv,~P.; Wijshoff,~H.; Versluis,~M.; Lohse,~D.
  Gravitational effect in evaporating binary microdroplets. \emph{Phys. Rev.
  Lett.} \textbf{2019}, \emph{122}, 114501\relax
\mciteBstWouldAddEndPuncttrue
\mciteSetBstMidEndSepPunct{\mcitedefaultmidpunct}
{\mcitedefaultendpunct}{\mcitedefaultseppunct}\relax
\EndOfBibitem
\bibitem[Diddens \latin{et~al.}(2017)Diddens, Tan, Lv, Versluis, Kuerten,
  Zhang, and Lohse]{diddens2017evaporating}
Diddens,~C.; Tan,~H.; Lv,~P.; Versluis,~M.; Kuerten,~J.; Zhang,~X.; Lohse,~D.
  Evaporating pure, binary and ternary droplets: thermal effects and axial
  symmetry breaking. \emph{J. Fluid Mech.} \textbf{2017}, \emph{823},
  470--497\relax
\mciteBstWouldAddEndPuncttrue
\mciteSetBstMidEndSepPunct{\mcitedefaultmidpunct}
{\mcitedefaultendpunct}{\mcitedefaultseppunct}\relax
\EndOfBibitem
\bibitem[Sefiane \latin{et~al.}(2003)Sefiane, Tadrist, and
  Douglas]{sefiane2003experimental}
Sefiane,~K.; Tadrist,~L.; Douglas,~M. Experimental study of evaporating
  water--ethanol mixture sessile drop: influence of concentration. \emph{Int.
  J. Heat Mass Transf.} \textbf{2003}, \emph{46}, 4527--4534\relax
\mciteBstWouldAddEndPuncttrue
\mciteSetBstMidEndSepPunct{\mcitedefaultmidpunct}
{\mcitedefaultendpunct}{\mcitedefaultseppunct}\relax
\EndOfBibitem
\bibitem[Sefiane \latin{et~al.}(2008)Sefiane, David, and
  Shanahan]{sefiane2008wetting}
Sefiane,~K.; David,~S.; Shanahan,~M. Wetting and evaporation of binary mixture
  drops. \emph{J. Phys. Chem. B.} \textbf{2008}, \emph{112}, 11317--11323\relax
\mciteBstWouldAddEndPuncttrue
\mciteSetBstMidEndSepPunct{\mcitedefaultmidpunct}
{\mcitedefaultendpunct}{\mcitedefaultseppunct}\relax
\EndOfBibitem
\bibitem[Yonemoto \latin{et~al.}(2018)Yonemoto, Suzuki, Uenomachi, and
  Kunugi]{yonemoto2018sliding}
Yonemoto,~Y.; Suzuki,~S.; Uenomachi,~S.; Kunugi,~T. Sliding behaviour of
  water-ethanol mixture droplets on inclined low-surface-energy solid.
  \emph{Int. J. Heat Mass Transf.} \textbf{2018}, \emph{120}, 1315--1324\relax
\mciteBstWouldAddEndPuncttrue
\mciteSetBstMidEndSepPunct{\mcitedefaultmidpunct}
{\mcitedefaultendpunct}{\mcitedefaultseppunct}\relax
\EndOfBibitem
\bibitem[Edwards \latin{et~al.}(2018)Edwards, Atkinson, Cheung, Liang,
  Fairhurst, and Ouali]{edwards2018density}
Edwards,~A.; Atkinson,~P.; Cheung,~C.; Liang,~H.; Fairhurst,~D.; Ouali,~F.
  Density-driven flows in evaporating binary liquid droplets. \emph{Phys. Rev.
  Lett.} \textbf{2018}, \emph{121}, 184501\relax
\mciteBstWouldAddEndPuncttrue
\mciteSetBstMidEndSepPunct{\mcitedefaultmidpunct}
{\mcitedefaultendpunct}{\mcitedefaultseppunct}\relax
\EndOfBibitem
\bibitem[Mamalis \latin{et~al.}(2016)Mamalis, Koutsos, and
  Sefiane]{mamalis2016motion}
Mamalis,~D.; Koutsos,~V.; Sefiane,~K. On the motion of a sessile drop on an
  incline: Effect of non-monotonic thermocapillary stresses. \emph{Appl. Phys.
  Lett.} \textbf{2016}, \emph{109}, 231601\relax
\mciteBstWouldAddEndPuncttrue
\mciteSetBstMidEndSepPunct{\mcitedefaultmidpunct}
{\mcitedefaultendpunct}{\mcitedefaultseppunct}\relax
\EndOfBibitem
\bibitem[Katre \latin{et~al.}(2021)Katre, Balusamy, Banerjee, Chandrala, and
  Sahu]{katre2021evaporation}
Katre,~P.; Balusamy,~S.; Banerjee,~S.; Chandrala,~L.~D.; Sahu,~K.~C.
  Evaporation dynamics of a sessile droplet of binary mixture laden with
  nanoparticles. \emph{Langmuir} \textbf{2021}, \emph{37}, 6311--6321\relax
\mciteBstWouldAddEndPuncttrue
\mciteSetBstMidEndSepPunct{\mcitedefaultmidpunct}
{\mcitedefaultendpunct}{\mcitedefaultseppunct}\relax
\EndOfBibitem
\bibitem[Katre \latin{et~al.}(2022)Katre, Balusamy, Banerjee, and
  Sahu]{katre2022experimental}
Katre,~P.; Balusamy,~S.; Banerjee,~S.; Sahu,~K.~C. An experimental
  investigation of evaporation of ethanol--water droplets laden with alumina
  nanoparticles on a critically inclined heated substrate. \emph{Langmuir}
  \textbf{2022}, \emph{38}, 4722--4735\relax
\mciteBstWouldAddEndPuncttrue
\mciteSetBstMidEndSepPunct{\mcitedefaultmidpunct}
{\mcitedefaultendpunct}{\mcitedefaultseppunct}\relax
\EndOfBibitem
\bibitem[Gurrala \latin{et~al.}(2019)Gurrala, Katre, Balusamy, Banerjee, and
  Sahu]{gurrala2019evaporation}
Gurrala,~P.; Katre,~P.; Balusamy,~S.; Banerjee,~S.; Sahu,~K.~C. Evaporation of
  ethanol-water sessile droplet of different compositions at an elevated
  substrate temperature. \emph{Int. J. Heat Mass Transf.} \textbf{2019},
  \emph{145}, 118770\relax
\mciteBstWouldAddEndPuncttrue
\mciteSetBstMidEndSepPunct{\mcitedefaultmidpunct}
{\mcitedefaultendpunct}{\mcitedefaultseppunct}\relax
\EndOfBibitem
\bibitem[Olayiwola and Dejam(2019)Olayiwola, and
  Dejam]{olayiwola2019mathematical}
Olayiwola,~S.~O.; Dejam,~M. Mathematical modelling of surface tension of
  nanoparticles in electrolyte solutions. \emph{Chem. Eng. Sci.} \textbf{2019},
  \emph{197}, 345--356\relax
\mciteBstWouldAddEndPuncttrue
\mciteSetBstMidEndSepPunct{\mcitedefaultmidpunct}
{\mcitedefaultendpunct}{\mcitedefaultseppunct}\relax
\EndOfBibitem
\bibitem[Vazquez \latin{et~al.}(1995)Vazquez, Alvarez, and
  Navaza]{vazquez1995surface}
Vazquez,~G.; Alvarez,~E.; Navaza,~J.~M. Surface tension of alcohol water+ water
  from 20 to 50 degree {C}. \emph{J. Chem. Eng. Data} \textbf{1995}, \emph{40},
  611--614\relax
\mciteBstWouldAddEndPuncttrue
\mciteSetBstMidEndSepPunct{\mcitedefaultmidpunct}
{\mcitedefaultendpunct}{\mcitedefaultseppunct}\relax
\EndOfBibitem
\bibitem[Tanvir and Qiao(2012)Tanvir, and Qiao]{tanvir2012surface}
Tanvir,~S.; Qiao,~L. Surface tension of nanofluid-type fuels containing
  suspended nanomaterials. \emph{Nanoscale Res. Lett.} \textbf{2012}, \emph{7},
  1--10\relax
\mciteBstWouldAddEndPuncttrue
\mciteSetBstMidEndSepPunct{\mcitedefaultmidpunct}
{\mcitedefaultendpunct}{\mcitedefaultseppunct}\relax
\EndOfBibitem
\bibitem[Macdougall and Ockrent(1942)Macdougall, and
  Ockrent]{macdougall1942surface}
Macdougall,~G.; Ockrent,~C. Surface energy relations in liquid/solid systems I.
  The adhesion of liquids to solids and a new method of determining the surface
  tension of liquids. \emph{Proc. Math. Phys. Eng. Sci.} \textbf{1942},
  \emph{180}, 151--173\relax
\mciteBstWouldAddEndPuncttrue
\mciteSetBstMidEndSepPunct{\mcitedefaultmidpunct}
{\mcitedefaultendpunct}{\mcitedefaultseppunct}\relax
\EndOfBibitem
\bibitem[Goodwin \latin{et~al.}(1988)Goodwin, Rice, and
  Middleman]{goodwin1988model}
Goodwin,~R.; Rice,~D.; Middleman,~S. A model for the onset of motion of a
  sessile liquid drop on a rotating disk. \emph{J. Colloid Interf. Sci.}
  \textbf{1988}, \emph{125}, 162--169\relax
\mciteBstWouldAddEndPuncttrue
\mciteSetBstMidEndSepPunct{\mcitedefaultmidpunct}
{\mcitedefaultendpunct}{\mcitedefaultseppunct}\relax
\EndOfBibitem
\bibitem[Qu{\'e}r{\'e} \latin{et~al.}(1998)Qu{\'e}r{\'e}, Azzopardi, and
  Delattre]{quere1998drops}
Qu{\'e}r{\'e},~D.; Azzopardi,~M.-J.; Delattre,~L. Drops at rest on a tilted
  plane. \emph{Langmuir} \textbf{1998}, \emph{14}, 2213--2216\relax
\mciteBstWouldAddEndPuncttrue
\mciteSetBstMidEndSepPunct{\mcitedefaultmidpunct}
{\mcitedefaultendpunct}{\mcitedefaultseppunct}\relax
\EndOfBibitem
\bibitem[Extrand and Kumagai(1997)Extrand, and
  Kumagai]{extrand1997experimental}
Extrand,~C.~W.; Kumagai,~Y. An experimental study of contact angle hysteresis.
  \emph{J. Colloid Interf. Sci.} \textbf{1997}, \emph{191}, 378--383\relax
\mciteBstWouldAddEndPuncttrue
\mciteSetBstMidEndSepPunct{\mcitedefaultmidpunct}
{\mcitedefaultendpunct}{\mcitedefaultseppunct}\relax
\EndOfBibitem
\bibitem[Chibowski \latin{et~al.}(2002)Chibowski, Ontiveros-Ortega, and
  Perea-Carpio]{chibowski2002interpretation}
Chibowski,~E.; Ontiveros-Ortega,~A.; Perea-Carpio,~R. On the interpretation of
  contact angle hysteresis. \emph{J. Adhes. Sci. Technol.} \textbf{2002},
  \emph{16}, 1367--1404\relax
\mciteBstWouldAddEndPuncttrue
\mciteSetBstMidEndSepPunct{\mcitedefaultmidpunct}
{\mcitedefaultendpunct}{\mcitedefaultseppunct}\relax
\EndOfBibitem
\bibitem[Extrand(2002)]{extrand2002water}
Extrand,~C. Water contact angles and hysteresis of polyamide surfaces. \emph{J.
  Colloid Interf. Sci.} \textbf{2002}, \emph{248}, 136--142\relax
\mciteBstWouldAddEndPuncttrue
\mciteSetBstMidEndSepPunct{\mcitedefaultmidpunct}
{\mcitedefaultendpunct}{\mcitedefaultseppunct}\relax
\EndOfBibitem
\bibitem[Schmidt \latin{et~al.}(2004)Schmidt, Brady, Lam, Schmidt, and
  Chaudhury]{schmidt2004contact}
Schmidt,~D.~L.; Brady,~R.~F.; Lam,~K.; Schmidt,~D.~C.; Chaudhury,~M.~K. Contact
  angle hysteresis, adhesion, and marine biofouling. \emph{Langmuir}
  \textbf{2004}, \emph{20}, 2830--2836\relax
\mciteBstWouldAddEndPuncttrue
\mciteSetBstMidEndSepPunct{\mcitedefaultmidpunct}
{\mcitedefaultendpunct}{\mcitedefaultseppunct}\relax
\EndOfBibitem
\bibitem[Faibish \latin{et~al.}(2002)Faibish, Yoshida, and
  Cohen]{faibish2002contact}
Faibish,~R.~S.; Yoshida,~W.; Cohen,~Y. Contact angle study on polymer-grafted
  silicon wafers. \emph{J. Colloid Interf. Sci.} \textbf{2002}, \emph{256},
  341--350\relax
\mciteBstWouldAddEndPuncttrue
\mciteSetBstMidEndSepPunct{\mcitedefaultmidpunct}
{\mcitedefaultendpunct}{\mcitedefaultseppunct}\relax
\EndOfBibitem
\end{mcitethebibliography}

\providecommand{\latin}[1]{#1}
\makeatletter
\providecommand{\doi}
  {\begingroup\let\do\@makeother\dospecials
  \catcode`\{=1 \catcode`\}=2 \doi@aux}
\providecommand{\doi@aux}[1]{\endgroup\texttt{#1}}
\makeatother
\providecommand*\mcitethebibliography{\thebibliography}
\csname @ifundefined\endcsname{endmcitethebibliography}
  {\let\endmcitethebibliography\endthebibliography}{}

 \end{document}